\UseRawInputEncoding
\documentclass[11pt]{article}
\pdfoutput=1
\usepackage{amsmath,amssymb,amsthm,amsxtra,overpic,bbm,bm,epsfig,ulem,color,multirow}
\begin{document}

\begin{center}
{\Large \bf Renormalization group evolution induced leptogenesis \\
in the minimal seesaw model \\ \vspace{0.2cm}
with the trimaximal mixing and mu-tau reflection symmetry }
\end{center}

\vspace{0.05cm}

\begin{center}
{\bf Zhen-hua Zhao\footnote{E-mail: zhaozhenhua@lnnu.edu.cn}} \\
{ Department of Physics, Liaoning Normal University, Dalian 116029, China }
\end{center}

\vspace{0.2cm}

\begin{abstract}
In this paper, we consider the imbedding of the popular and well-motivated trimaximal mixing and $\mu$-$\tau$ reflection symmetry (which can help us shape the forms of the neutrino mass matrix) in the minimal seesaw model (which contains much fewer parameters than the general seesaw model) with two TeV-scale right-handed neutrinos (for realizing a low-scale seesaw) of nearly degenerate masses (for realizing a resonant leptogenesis). However, either for the trimaximal mixing scenario (which is realized through the Form Dominance approach here) or for the $\mu$-$\tau$ reflection symmetry scenario, leptogenesis cannot proceed. To address this issue, we consider the possibility that the special forms of the neutrino mass matrix for the trimaximal mixing and $\mu$-$\tau$ reflection symmetry are slightly broken by the renormalization group evolution effect, thus allowing leptogenesis to proceed. It is found that in the normal case of the neutrino mass ordering, the baryon asymmetry thus generated can reproduce the observed value. For completeness, we have also extended our analysis to the scenario that two right-handed neutrinos are not nearly degenerate any more. Unfortunately, in this scenario the final baryon asymmetry is smaller than the observed value by several orders of magnitude.
\end{abstract}

\newpage

\section{Introduction}

As we know, the phenomena of neutrino oscillations show that neutrinos are massive and the lepton flavors are mixed \cite{xing}. In the literature, the most popular and natural way of generating the tiny but non-zero neutrino masses is the type-I seesaw mechanism in which three heavy right-handed neutrino fields $N^{}_I$ (for $I=1, 2, 3$) are introduced into the SM \cite{seesaw}. They not only have Yukawa couplings $(Y^{}_\nu)^{}_{\alpha I}$ with the left-handed lepton doublets $L^{}_{\alpha}$ (for $\alpha = e, \mu, \tau$), which yield the Dirac neutrino mass matrix $(M^{}_{\rm D})^{}_{\alpha I} = (Y^{}_\nu)^{}_{\alpha I} v$ after the Higgs field acquires its vacumm expectation value (VEV) $v = 174$ GeV, but themselves also constitute a Majorana mass matrix $M^{}_{\rm R}$. Without loss of generality, we will work in the basis of $M^{}_{\rm R}$ being diagonal $D^{}_{\rm R} = {\rm diag}(M^{}_1, M^{}_2, M^{}_3)$ with $M^{}_I$ being three right-handed neutrino masses.
After integrating out these right-handed neutrino fields, an effective Majorana mass matrix for the light neutrinos arises as
$M^{}_{\nu} \simeq - M^{}_{\rm D} M^{-1}_{\rm R} M^{T}_{\rm D}$.
 For the following reason, the seesaw scale (i.e., the right-handed neutrino mass scale) is usually taken to be extremely high: if the neutrino Yukawa couplings take ${\cal{O}}(1)$ values, then the sub-eV light neutrino masses will be achieved in correspondence to ${\cal{O}}(10^{14})$ GeV right-handed neutrino masses via the seesaw formula.

In the basis where the flavor eigenstates of three charged leptons align with their mass eigenstates (i.e., $Y^{}_l = {\rm diag}(y^{}_e, y^{}_\mu, y^{}_\tau)$ with $y^{}_\alpha = m^{}_\alpha/v$), the neutrino mixing matrix $U$ \cite{pmns} is identical with the unitary matrix for diagonalizing $M^{}_\nu$: $U^\dagger M^{}_\nu U^* =  D^{}_\nu =  {\rm diag}(m^{}_1, m^{}_2, m^{}_3) $
with $m^{}_i$ being three light neutrino masses.
In the standard parametrization, $U$ is expressed in terms of three mixing angles $\theta^{}_{ij}$ (for $ij=12, 13, 23$), one Dirac CP phase $\delta$, two Majorana CP phases $\rho$ and $\sigma$, and three unphysical phases $\phi^{}_\alpha$
\begin{eqnarray}
U  =
\left( \begin{matrix}
e^{{\rm i}\phi^{}_e} &  & \cr
& e^{{\rm i}\phi^{}_\mu}  & \cr
&  & e^{{\rm i}\phi^{}_\tau}
\end{matrix} \right) \left( \begin{matrix}
c^{}_{12} c^{}_{13} & s^{}_{12} c^{}_{13} & s^{}_{13} e^{-{\rm i} \delta} \cr
-s^{}_{12} c^{}_{23} - c^{}_{12} s^{}_{23} s^{}_{13} e^{{\rm i} \delta}
& c^{}_{12} c^{}_{23} - s^{}_{12} s^{}_{23} s^{}_{13} e^{{\rm i} \delta}  & s^{}_{23} c^{}_{13} \cr
s^{}_{12} s^{}_{23} - c^{}_{12} c^{}_{23} s^{}_{13} e^{{\rm i} \delta}
& -c^{}_{12} s^{}_{23} - s^{}_{12} c^{}_{23} s^{}_{13} e^{{\rm i} \delta} & c^{}_{23}c^{}_{13}
\end{matrix} \right) \left( \begin{matrix}
e^{{\rm i}\rho} &  & \cr
& e^{{\rm i}\sigma}  & \cr
&  & 1
\end{matrix} \right) \;,
\label{1}
\end{eqnarray}
where the abbreviations $c^{}_{ij} = \cos \theta^{}_{ij}$ and $s^{}_{ij} = \sin \theta^{}_{ij}$ have been employed.

Neutrino oscillations are sensitive to three mixing angles, the neutrino mass squared differences $\Delta m^2_{ij} \equiv m^2_i - m^2_j$, and $\delta$. Several groups have performed global analyses of the existing neutrino oscillation data to extract the values of these parameters \cite{global,global2}. For definiteness, we will use the results in Ref.~\cite{global} (see Table~1) as reference values in the following numerical calculations. Since the sign of $\Delta m^2_{31}$ remains undetermined, there are two possibilities for the neutrino mass ordering: the normal ordering (NO) $m^{}_1 < m^{}_2 < m^{}_3$ and inverted ordering (IO) $m^{}_3 < m^{}_1 < m^{}_2$. But neutrino oscillations have nothing to do with the absolute neutrino masses and Majorana CP phases, whose values can only be inferred from non-oscillatory experiments. So far, there has not yet been any lower constraint on the lightest neutrino mass, nor any constraint on the Majorana CP phases.

It has long been noticed that $\theta^{}_{12}$ and $\theta^{}_{23}$ are close to some special values: $\sin^2 \theta^{}_{12} \sim 1/3$ and $\sin^2 \theta^{}_{23} \sim 1/2$. And $\theta^{}_{13}$ seemed likely to be vanishingly small before its value was pinned down. For the ideal case of $\sin \theta^{}_{12} = 1/\sqrt{3}$, $\sin \theta^{}_{23} = 1/\sqrt{2}$ and $\theta^{}_{13} =0$ (referred to as the tribimaximal (TBM) mixing \cite{TB}), the neutrino mixing matrix can be described by some simple numbers and their square roots:
\begin{eqnarray}
U^{}_{\rm TBM}= \displaystyle \frac{1}{\sqrt 6} \left( \begin{array}{ccc} \vspace{0.15cm}
2 & \sqrt{2} & 0 \cr \vspace{0.15cm}
- 1 & \sqrt{2}  & \sqrt{3}  \cr
1 & - \sqrt{2}  & \sqrt{3} \cr
\end{array} \right)  \; .
\label{2}
\end{eqnarray}
Such a special mixing pattern is suggestive of some underlying flavor symmetry in the lepton sector.
In the literature many flavor symmetries have been tentatively employed to realize it \cite{FS}. However, the relative largeness of $\theta^{}_{13}$ compels us to forsake or modify this simple but instructive mixing pattern. An economical and predictive way out is to retain its first or second column while modifying the other two columns within the unitarity constraints, giving the first or second trimaximal (TM1 or TM2) mixing \cite{TM}
\begin{eqnarray}
U^{}_{\rm TM1}=  \displaystyle \frac{1}{\sqrt 6} \left( \begin{array}{ccc} \vspace{0.15cm}
2 & \cdot & \cdot \cr \vspace{0.15cm}
- 1 & \cdot & \cdot \cr
1 & \cdot & \cdot \cr
\end{array} \right)  \;, \hspace{1cm}
U^{}_{\rm TM2}=  \displaystyle \frac{1}{\sqrt 3} \left( \begin{array}{ccc} \vspace{0.15cm}
\cdot & 1 & \cdot \cr \vspace{0.15cm}
\cdot & 1 &  \cdot \cr
\cdot &  -1 & \cdot \cr
\end{array} \right)  \; .
\label{3}
\end{eqnarray}

As is known, the flavor symmetry corresponding to $\sin \theta^{}_{23} = 1/\sqrt{2}$ and $\theta^{}_{13} =0$ is the $\mu$-$\tau$ symmetry, which requires the neutrino mass matrix to keep invariant with respect to the interchange between the $\mu$ and $\tau$ flavor left-handed neutrino fields (i.e., $\nu^{}_{\mu} \leftrightarrow \nu^{}_\tau$) \cite{mutau1, mutau2}. After the observation of a relatively large $\theta^{}_{13}$ and a preliminary hint for $\delta \sim - \pi/2$ \cite{T2K}, the $\mu$-$\tau$ reflection symmetry \cite{mu-tauR, mutau2} --- a generalized $\mu$-$\tau$ symmetry --- has become increasingly popular, which requires the neutrino mass matrix to keep invariant with respect to the following transformations of three left-handed neutrino fields
\begin{eqnarray}
\nu^{}_{e} \leftrightarrow \nu^{c}_e \;, \hspace{1cm} \nu^{}_{\mu} \leftrightarrow \nu^{c}_{\tau} \;,
\hspace{1cm} \nu^{}_{\tau} \leftrightarrow \nu^{c}_{\mu} \;,
\label{4}
\end{eqnarray}
where the superscript ¡®$c$¡¯ denotes the charge conjugation of relevant neutrino fields.
Such a symmetry leads to the following interesting predictions for the neutrino mixing parameters
\begin{eqnarray}
\sin \theta^{}_{23} = \frac{1}{\sqrt{2}} \;, \hspace{1cm} \delta = \pm \frac{\pi}{2} \;,
\hspace{1cm} \rho, \sigma = 0 \ {\rm or} \ \frac{\pi}{2} \;, \hspace{1cm} \phi^{}_e = \frac{\pi}{2}  \;, \hspace{1cm} \phi^{}_\mu = - \phi^{}_\tau  \;.
\label{5}
\end{eqnarray}
In the following calculations relevant for the $\mu$-$\tau$ reflection symmetry, we will fix $\delta$ to be $-\pi/2$ which is more favored experimentally.

As an extra bonus, the seesaw mechanism also provides an appealing explanation for the baryon asymmetry of the Universe \cite{planck}
\begin{eqnarray}
Y^{0}_{\rm B} \equiv \frac{n^{}_{\rm B}-n^{}_{\rm \bar B}}{s^{}_0} \simeq (8.69 \pm 0.04) \times 10^{-11}  \;,
\label{6}
\end{eqnarray}
where $s^{}_0$ denotes the entropy density in the present epoch. This is just the leptogenesis mechanism \cite{leptogenesis, Lreview}: a lepton asymmetry is firstly generated by the lepton-number-violating and CP-violating decays of right-handed neutrinos and then partially converted into the baryon asymmetry via the sphaleron processes \cite{sphaleron}. The requirement of successful leptogenesis places a lower bound $\sim 10^{9}$ GeV for the right-handed neutrino masses \cite{DI} which translates to a comparable lower bound for the reheating temperature of the Universe after the inflationary period. Note that such a high reheating temperature will be problematic for supersymmetric theories because it leads to an overproduction of light states such as the gravitino which would spoil the successful predictions of the big bang nucleosyntheis \cite{gravitino}.

Despite its remarkable merits, the conventional type-I seesaw model with three super heavy right-handed neutrino fields have two apparent shortcomings:
(1) the number of its parameters is significantly larger than that of the low-energy neutrino parameters, rendering it incapable of giving any quantitative prediction for the latter or leptogenesis. In order to improve its predictability, there are two typical approaches to reducing its parameters. One approach is to constrain its flavor structure by employing some flavor symmetry such as the aforementioned trimaximal symmetry (i.e., the flavor symmetry associated with the trimaximal mixing) and $\mu$-$\tau$ reflection symmetry. The other approach is to reduce the number of right-handed neutrino fields to two (i.e., the minimal seesaw model \cite{MSS, MSS2}), in which case the lightest neutrino mass remains to be vanishing (i.e., $m^{}_1=0$ in the NO case or $m^{}_3 =0$ in the IO case) and only one Majorana CP phase is physically relevant ($\sigma$ in the NO case or $\sigma -\rho$ in the IO case). (2) The conventional seesaw scale is too high to be accessed by current experiments. As a matter of fact, the Yukawa couplings of different fermions span many orders of magnitude, from ${\cal{O}}(10^{-6})$ (the electron) to $\simeq 1$ (the top quark).
If the neutrino Yukawa couplings are somewhat comparable with the electron Yukawa coupling, then the sub-eV light neutrino masses will be achieved in correspondence to only TeV-scale right-handed neutrino masses via the seesaw formula. In this sense a low-scale seesaw model is absolutely acceptable \cite{lowscale}. Furthermore, when the right-handed neutrinos are nearly degenerate, leptogenesis will get resonantly enhanced so that a successful leptogenesis can also be fulfilled for low right-handed neutrino masses (evading the aforementioned lower bound for the right-handed neutrino masses) \cite{resonant}.

In view of the above facts, it will be an interesting attempt to embed the popular and well-motivated trimaximal mixing and $\mu$-$\tau$ reflection symmetry (for constraining the flavor structure) in the minimal seesaw model (for reducing the model parameters) with two TeV-scale right-handed neutrinos (for realizing a low-scale seesaw) of nearly degenerate masses (for realizing a resonant leptogenesis). However, as will be seen, leptogenesis dose not work either in the trimaximal mixing scenario (which is realized through the Form Dominance approach here) or in the $\mu$-$\tau$ reflection symmetry scenario.
Fortunately, the renormalization group evolution (RGE) effect is potentially capable of inducing a viable leptogenesis for both of these two scenarios.
In sections 3 and 4 we study such an interesting possibility for the trimaximal mixing and $\mu$-$\tau$ reflection symmetry scenarios, respectively. In section 5 we extend our analysis to the scenario that two right-handed neutrinos are not nearly degenerate any more.

\begin{table}\centering
  \begin{footnotesize}
    \begin{tabular}{c|cc|cc}
     \hline\hline
      & \multicolumn{2}{c|}{Normal Ordering}
      & \multicolumn{2}{c}{Inverted Ordering }
      \\
      \cline{2-5}
      & bf $\pm 1\sigma$ & $3\sigma$ range
      & bf $\pm 1\sigma$ & $3\sigma$ range
      \\
      \cline{1-5}
      \rule{0pt}{4mm}\ignorespaces
       $\sin^2\theta^{}_{12}$
      & $0.318_{-0.016}^{+0.016}$ & $0.271 \to 0.370$
      & $0.318_{-0.016}^{+0.016}$ & $0.271 \to 0.370$
      \\[1mm]
       $\sin^2\theta^{}_{23}$
      & $0.566_{-0.022}^{+0.016}$ & $0.441 \to 0.609$
      & $0.566_{-0.023}^{+0.018}$ & $0.446 \to 0.609$
      \\[1mm]
       $\sin^2\theta^{}_{13}$
      & $0.02225_{-0.00078}^{+0.00055}$ & $0.02015 \to 0.02417$
      & $0.02250_{-0.00076}^{+0.00056}$ & $0.02039 \to 0.02441$
      \\[1mm]
       $\delta/\pi$
      & $1.20_{-0.14}^{+0.23}$ & $0.80 \to 2.00$
      & $1.54_{-0.13}^{+0.13}$ & $1.14 \to 1.90$
      \\[3mm]
       $\Delta m^2_{21}/(10^{-5}~{\rm eV}^2)$
      & $7.50_{-0.20}^{+0.22}$ & $6.94 \to 8.14$
      & $7.50_{-0.20}^{+0.22}$ & $6.94 \to 8.14$
      \\[3mm]
       $|\Delta m^2_{31}|/(10^{-3}~{\rm eV}^2)$
      & $2.56_{-0.04}^{+0.03}$ & $2.46 \to 2.65$
      & $2.46_{-0.03}^{+0.03}$ & $2.37 \to 2.55$
      \\[2mm]
      \hline\hline
    \end{tabular}
  \end{footnotesize}
  \caption{The best-fit values, 1$\sigma$ errors and 3$\sigma$ ranges of six neutrino
oscillation parameters extracted from a global analysis of the existing
neutrino oscillation data \cite{global}. }
\end{table}

\section{Some basics}

Before performing our study, we first give some basics for leptogenesis and RGEs of the neutrino mass matrices.

\subsection{Some basics for leptogenesis}

It is well known that, according to the temperature where leptogenesis takes place (the right-handed neutrino mass scale), there are three possible regimes for leptogenesis \cite{flavor}. (1) Unflavored regime: in the temperature range above $10^{12}$ GeV where the charged-lepton Yukawa $y^{}_\alpha$ interactions have not yet entered thermal equilibrium, three lepton flavors are indistinguishable and should be treated in a universal manner. (2) Two-flavor regime: in the temperature range $10^{9}$---$10^{12}$ GeV where the $y^{}_\tau$-related interactions are in thermal equilibrium, the $\tau$ flavor is distinguishable from the other two flavors which remain indistinguishable and should be treated separately. (3) Three-flavor regime: in the temperature range below $10^{9}$ GeV where the $y^{}_\mu$-related interactions also enter thermal equilibrium, all the three flavors are distinguishable and should be treated separately.

We will perform a study for the following two scenarios separately. (1) For the low-scale (leading us to the three-flavor regime) resonant leptogenesis scenario, the final baryon asymmetry is given by \cite{resonant}
\begin{eqnarray}
Y^{}_{\rm B} = c r \sum^{}_{\alpha} \varepsilon^{}_{\alpha}  \kappa ( \widetilde m^{}_\alpha) = c r \left[ \varepsilon^{}_{e}  \kappa ( \widetilde m^{}_e ) +
\varepsilon^{}_{\mu}  \kappa ( \widetilde m^{}_\mu ) + \varepsilon^{}_{\tau}  \kappa ( \widetilde m^{}_\tau )  \right] \;,
\label{2.1.1}
\end{eqnarray}
where $c = - 28/79$ describes the transition efficiency from the lepton asymmetry to the baryon asymmetry via the sphaleron processes, and $r \simeq 4 \times 10^{-3}$ measures the ratio of the equilibrium $N^{}_I$ number density to the entropy density at the temperature above $M^{}_I$. $\varepsilon^{}_{\alpha}$ is the sum of the flavored CP asymmetries $\varepsilon^{}_{I\alpha}$ over $I$ (i.e., $\varepsilon^{}_{\alpha} = \varepsilon^{}_{1\alpha} + \varepsilon^{}_{2\alpha}$), which quantify the asymmetries between the decay rates of $N^{}_I \to L^{}_\alpha + H$ and their CP-conjugate processes $N^{}_I \to \overline{L}^{}_\alpha + \overline{H}$. And $\varepsilon^{}_{I\alpha}$ are explicitly given by
\cite{resonant}
\begin{eqnarray}
\varepsilon^{}_{I\alpha} = \frac{{\rm Im}\left\{ (M^*_{\rm D})^{}_{\alpha I} (M^{}_{\rm D})^{}_{\alpha J}
\left[ M^{}_J (M^\dagger_{\rm D} M^{}_{\rm D})^{}_{IJ} + M^{}_I (M^\dagger_{\rm D} M^{}_{\rm D})^{}_{JI} \right] \right\} }{8\pi  v^2 (M^\dagger_{\rm D} M^{}_{\rm D})^{}_{II}} \cdot \frac{M^{}_I \Delta M^2_{IJ}}{(\Delta M^2_{IJ})^2 + M^2_I \Gamma^2_J} \;,
\label{2.1.2}
\end{eqnarray}
where $\Delta M^2_{IJ} \equiv M^2_I - M^2_J$ has been defined and $\Gamma^{}_J= (M^\dagger_{\rm D} M^{}_{\rm D})^{}_{JJ} M^{}_J/(8\pi v^2)$ is the decay rate of $N^{}_J$ (for $J \neq I$). Finally, the efficiency factor $\kappa \left( \widetilde m^{}_\alpha\right) <1$ takes account of the washout effects due to the inverse decay and various lepton-number-violating scattering processes. Its value is determined by the washout mass parameter $\widetilde m^{}_\alpha = \widetilde m^{}_{1 \alpha} + \widetilde m^{}_{2 \alpha}$ with
$\widetilde m^{}_{I \alpha} = |(M^{}_{\rm D})^{}_{\alpha I}|^2/M^{}_I$. We see that in the present scenario two right-handed neutrinos are on an equal footing in contributing to leptogenesis. This is simply because they are nearly degenerate.

(2) But in the scenario that two right-handed neutrinos are not nearly degenerate any more, the contribution to leptogenesis mainly comes from the lighter one because that from the heavier one suffers from its washout effects. In the unflavored regime, the final baryon asymmetry is given by
\begin{eqnarray}
Y^{}_{\rm B} = c r \varepsilon^{}_I \kappa(\widetilde m^{}_I)  \;,
\label{2.1.3}
\end{eqnarray}
with $I$ denoting the lighter right-handed neutrino and $\widetilde m^{}_I = \widetilde m^{}_{I e} + \widetilde m^{}_{I \mu} + \widetilde m^{}_{I \tau}$. $\varepsilon^{}_I$ is the total CP asymmetry for the decays of $N^{}_I$
\begin{eqnarray}
\varepsilon^{}_{I} = \frac{1}{8\pi (M^\dagger_{\rm D}
M^{}_{\rm D})^{}_{II} v^2} {\rm Im}\left[ (M^\dagger_{\rm D} M^{}_{\rm D})^{2}_{IJ}\right] {\cal F} \left( \frac{M^2_J}{M^2_I} \right) \;,
\label{2.1.4}
\end{eqnarray}
with $J \neq I$ denoting the heavier right-handed neutrino, which is a sum of the flavored CP asymmetries
\begin{eqnarray}
\varepsilon^{}_{I \alpha} & = & \frac{1}{8\pi (M^\dagger_{\rm D}
M^{}_{\rm D})^{}_{II} v^2} \left\{ {\rm Im}\left[(M^*_{\rm D})^{}_{\alpha I} (M^{}_{\rm D})^{}_{\alpha J}
(M^\dagger_{\rm D} M^{}_{\rm D})^{}_{IJ}\right] {\cal F} \left( \frac{M^2_J}{M^2_I} \right) \right. \nonumber \\
&  &
+ \left. {\rm Im}\left[(M^*_{\rm D})^{}_{\alpha I} (M^{}_{\rm D})^{}_{\alpha J} (M^\dagger_{\rm D} M^{}_{\rm D})^*_{IJ}\right] {\cal G}  \left( \frac{M^2_J}{M^2_I} \right) \right\} \; ,
\label{2.1.5}
\end{eqnarray}
with ${\cal F}(x) = \sqrt{x} \{(2-x)/(1-x)+ (1+x) \ln [x/(1+x)] \}$ and ${\cal G}(x) = 1/(1-x)$.
In the two-flavor regime, the baryon asymmetry receives two contributions from $\varepsilon^{}_{I \tau}$ and $\varepsilon^{}_{I \gamma} = \varepsilon^{}_{I e} + \varepsilon^{}_{I \mu}$ which are subject to different washout effects controlled by $\widetilde m^{}_{I \tau}$ and $\widetilde m^{}_{I \gamma} = \widetilde m^{}_{I e} + \widetilde m^{}_{I \mu}$ \cite{flavor}
\begin{eqnarray}
Y^{}_{\rm B}
= c r \left[ \varepsilon^{}_{I \tau} \kappa \left(\frac{390}{589} \widetilde m^{}_{I \tau} \right) + \varepsilon^{}_{I \gamma} \kappa \left(\frac{417}{589} \widetilde m^{}_{I \gamma} \right) \right] \;.
\label{2.1.6}
\end{eqnarray}

\subsection{Some basics for RGEs of the neutrino mass matrices}

In the literature, the flavor symmetries that shape the special forms of the neutrino mass matrices are usually placed at a very high energy scale $\Lambda^{}_{\rm FS}$. When dealing with leptogenesis which takes place around the right-handed neutrino mass scale $M^{}_0$, one should take account of the renormalization group evolution effect if there is a large gap between $\Lambda^{}_{\rm FS}$ and $M^{}_0$ \cite{OZ}.

In the SM, at the one-loop level, the running behaviours of the Dirac neutrino mass matrix and right-handed neutrino mass matrix are described by \cite{ynu}
\begin{eqnarray}
&& 16 \pi^2 \frac{d M^{}_{\rm D}}{dt} = \left\{ \frac{3}{2} Y^{}_\nu Y^\dagger_\nu - \frac{3}{2} Y^{}_l Y^\dagger_l + {\rm Tr} \left[ 3 Y^{}_u Y^\dagger_u + 3 Y^{}_d Y^\dagger_d +  Y^{}_\nu Y^\dagger_\nu + Y^{}_l Y^\dagger_l \right] - \frac{9}{20} g^2_1 - \frac{9}{4} g^2_2 \right\}  M^{}_{\rm D} \;, \nonumber \\
&& 16 \pi^2 \frac{d M^{}_{\rm R}}{dt} = (Y^\dagger_\nu Y^{}_\nu)^T M^{}_{\rm R}  + M^{}_{\rm R} (Y^\dagger_\nu Y^{}_\nu) \;.
\label{2.2.1}
\end{eqnarray}
Here $t$ denotes $\ln(\mu/\Lambda^{}_{\rm FS})$ with $\mu$ being the renormalization scale, $Y^{}_{u, d}$ are the up-quark and down-quark Yukawa matrices and $g^{}_{1, 2}$ are the gauge couplings.

An integration of Eq.~(\ref{2.2.1}) enables us to obtain the Dirac neutrino mass matrix $M^{}_{\rm D}(M^{}_0)$ at the right-handed neutrino mass scale from its counterpart $M^{}_{\rm D}(\Lambda^{}_{\rm FS})$ at the flavor-symmetry scale  as \cite{RGE,IRGE}
\begin{eqnarray}
M^{}_{\rm D} (M^{}_0) = I^{}_{0} \left( \begin{array}{ccc}
1+\Delta^{}_{e} &   &  \cr
 & 1 +\Delta^{}_{\mu} &  \cr
 &  &  1+\Delta^{}_{\tau} \cr
\end{array} \right)
M^{}_{\rm D} (\Lambda^{}_{\rm FS}) \;,
\label{2.2.2}
\end{eqnarray}
where
\begin{eqnarray}
&& I^{}_{0}  =  {\rm exp} \left( - \frac{1}{16 \pi^2} \int^{\ln (\Lambda^{}_{\rm FS}/M^{}_0)}_{0} \left\{ {\rm Tr} \left[ 3 Y^{}_u Y^\dagger_u + 3 Y^{}_d Y^\dagger_d +  Y^{}_\nu Y^\dagger_\nu + Y^{}_l Y^\dagger_l \right] - \frac{9}{20} g^2_1 - \frac{9}{4} g^2_2 \right\} \ {\rm dt} \right) \;, \nonumber \\
&& \Delta^{}_{\alpha}   =   \frac{3}{32 \pi^2}\int^{\ln (\Lambda^{}_{\rm FS}/M^{}_0)}_{0} y^2_{\alpha} \ {\rm dt} \simeq \frac{3}{32 \pi^2} y^2_{\alpha} \ln \left(\frac{\Lambda^{}_{\rm FS}}{M^{}_0} \right) \;.
\label{2.2.3}
\end{eqnarray}
We see that $I^{}_0$ is just an overall rescaling factor. But $\Delta^{}_\alpha$ can modify the structure of $M^{}_{\rm D}$, due to their (i.e., $y^{}_{\alpha}$) differences. In the light of $\Delta^{}_{e} \ll \Delta^{}_{\mu} \ll \Delta^{}_{\tau}$ (as a result of $y^{}_e \ll y^{}_\mu \ll y^{}_\tau$), it is an excellent approximation for us to only keep $\Delta^{}_{\tau}$ in the following calculations.

On the other hand, the right-handed neutrino mass matrix $M^{}_{\rm R}(M^{}_0)$ at the right-handed neutrino mass scale is obtained from its counterpart $M^{}_{\rm R}(\Lambda^{}_{\rm FS}) = {\rm diag}(M^{0}_1, M^{0}_2)$ at the flavor-symmetry scale as
{\footnotesize
\begin{eqnarray}
M^{}_{\rm R} (M^{}_0) \simeq
\left( \begin{array}{cc}
M^{0}_1 - \displaystyle \frac{M^{0}_1}{8\pi^2} (Y^\dagger_\nu Y^{}_\nu)^{}_{11} \ln\left( \frac{\Lambda^{}_{\rm FS}}{M^{}_0} \right)  &  - \displaystyle \frac{1}{16\pi^2} \left\{ 2 M^0_1 {\rm Re}[(Y^\dagger_\nu Y^{}_\nu)^{}_{12}]  +  \Delta M^0_{21} (Y^\dagger_\nu Y^{}_\nu)^{*}_{12}  \right\} \ln\left( \frac{\Lambda^{}_{\rm FS}}{M^{}_0} \right)   \cr
\times & M^{0}_2 - \displaystyle \frac{M^{0}_2}{8\pi^2} (Y^\dagger_\nu Y^{}_\nu)^{}_{22} \ln\left( \frac{\Lambda^{}_{\rm FS}}{M^{}_0} \right)  \cr
\end{array} \right) \;,
\label{2.2.4}
\end{eqnarray}  }
with $\Delta M^0_{21} \equiv M^0_2 - M^0_1$ being the initial right-handed neutrino mass difference. If the off-diagonal terms of $M^{}_{\rm R} (M^{}_0)$ are vanishing, then the renormalization group evolution effect on the right-handed neutrino mass matrix is simply to add a new contribution to the right-handed neutrino mass difference:
\begin{eqnarray}
\Delta M \equiv M^{}_2 - M^{}_1 \simeq \Delta M^0_{21} + \displaystyle \frac{1}{8\pi^2} \left[ M^0_1 (Y^\dagger_\nu Y^{}_\nu)^{}_{11} - M^0_2 (Y^\dagger_\nu Y^{}_\nu)^{}_{22} \right] \ln\left( \frac{\Lambda^{}_{\rm FS}}{M^{}_0} \right) \;.
\label{2.2.5}
\end{eqnarray}
If the off-diagonal terms of $M^{}_{\rm R} (M^{}_0)$ are non-vanishing, then one can go back to the mass basis of the right-handed neutrinos via a unitary transformation $U^{}_{\rm R}$ of them: $U^{T}_{\rm R} M^{}_{\rm R} (M^{}_0) U^{}_{\rm R} = {\rm diag}(M^{}_1, M^{}_2)$. In the meantime, the Dirac neutrino mass matrix is transformed to $M^{\prime}_{\rm D} (M^{}_0) = M^{}_{\rm D} (M^{}_0) U^{}_{\rm R}$.

\section{Trimaximal mixing scenario}

\subsection{Trimaximal mixing in minimal seesaw}

In the literature, the idea of Form Dominance (FD) is a generic and natural mechanism that realizes a form diagonalizable $M^{}_\nu$ from which the resulting neutrino mixing matrix is independent of the parameters that control the neutrino masses \cite{form}. The basic idea of FD is that a specific right-handed neutrino is associated with a specific  light neutrino mass eigenstate. For example, if $M^{}_{\rm D}$ takes a form as (i.e., its three columns being respectively proportional to three columns of $U^{}_{\rm TBM}$)
\begin{eqnarray}
M^{}_{\rm D}= \left( \begin{array}{ccc} \vspace{0.15cm}
2 a \sqrt{M^{}_1} & b \sqrt{M^{}_2}  & 0 \cr \vspace{0.15cm}
- a \sqrt{M^{}_1} & b \sqrt{M^{}_2} & c \sqrt{M^{}_3} \cr
a \sqrt{M^{}_1} & - b \sqrt{M^{}_2} & c \sqrt{M^{}_3} \cr
\end{array} \right)  \;,
\label{3.1.1}
\end{eqnarray}
then the TBM mixing will arise automatically, independently of the parameters that control the neutrino masses
\begin{eqnarray}
m^{}_1 = 6 |a|^2 \;, \hspace{1cm} m^{}_2 = 3 |b|^2 \;, \hspace{1cm} m^{}_3 = 2 |c|^2 \;.
\label{3.1.2}
\end{eqnarray}
We see that three right-handed neutrinos are respectively associated with three light neutrino mass eigenstates.

In a concrete flavor-symmetry model, the form of $M^{}_{\rm D}$ in Eq.~(\ref{3.1.1}) is usually realized in a way as follows: under a properly chosen flavor symmetry (e.g., $A^{}_4$ and $S^{}_4$), three lepton doublet fields $L^{}_\alpha$ jointly constitute a triplet representation while three right-handed neutrino fields are simply singlet representations. Then one introduces three scalar flavon fields $\Phi^{}_{J}$ (for $J=1, 2, 3$), each of which is a triplet representation under the flavor symmetry. In this way the following kind of operators
\begin{eqnarray}
\sum^{}_{I, J} y^{}_{IJ}(\overline L^{}_e \Phi^{}_{J1} + \overline L^{}_\mu \Phi^{}_{J2} + \overline L^{}_\tau \Phi^{}_{J3} ) H N^{}_I \;,
\label{3.1.3}
\end{eqnarray}
which are singlet combinations under both the SM gauge and flavor symmetries, will serve to generate the Dirac neutrino masses after the Higgs and flavon fields acquire non-vanishing VEVs. Here $y^{}_{IJ}$ are some coefficients and $\Phi^{}_{J1}$ denotes the first component of $\Phi^{}_{J}$ (and so on). If three flavon fields are respectively associated with three right-handed neutrino fields (i.e., $y^{}_{IJ} = 0$ for $I \neq J$) and acquire the following particular VEV alignments
\begin{eqnarray}
\langle \Phi^{}_1 \rangle \propto (2, -1, 1)^T \;, \hspace{1cm} \langle \Phi^{}_2 \rangle \propto (1, 1, -1)^T \;, \hspace{1cm} \langle \Phi^{}_3 \rangle \propto (0, 1, 1)^T \;,
\label{3.1.4}
\end{eqnarray}
then the form of $M^{}_{\rm D}$ in Eq.~(\ref{3.1.1}) will be successfully reproduced.

\subsubsection{TM1 scenario}

Now, we consider the realization of trimaximal mixing in the minimal seesaw model. Following the idea of FD, one can achieve a TM1 mixing by having two columns of $M^{}_{\rm D}$ be respectively proportional and orthogonal to the first column of $U^{}_{\rm TBM}$ \cite{tminms}. Such an $M^{}_{\rm D}$ can be parameterized as
\begin{eqnarray}
M^{}_{\rm D}= \left( \begin{array}{cc}
\vspace{0.1cm}
2 a \sqrt{M^{}_1} &  b \sqrt{M^{}_2}  \cr
-a \sqrt{M^{}_1} & b (1-r) \sqrt{M^{}_2}  \cr
a \sqrt{M^{}_1} & -b (1+ r) \sqrt{M^{}_2}  \cr
\end{array} \right) \; ,
\label{3.1.5}
\end{eqnarray}
where $a$, $b$ and $r$ are generally complex parameters. It is easy to see that this form of  $M^{}_{\rm D}$ can be realized by associating  $\langle \Phi^{}_1 \rangle$ with $N^{}_1$ and both $\langle \Phi^{}_2 \rangle$ and $\langle \Phi^{}_3 \rangle$ with $N^{}_2$. And the dimensionless parameter $r$ measures the relative size between the contributions of $\langle \Phi^{}_2 \rangle$ and $\langle \Phi^{}_3 \rangle$ to the second column of $M^{}_{\rm D}$.

For the resulting neutrino mixing matrix, two columns (the remaining column) will be respectively proportional (orthogonal) to two columns of $M^{}_{\rm D}$ and associated with two non-vanishing (vanishing) light neutrino masses. Given that in the TM1 mixing the preserved first column of $U^{}_{\rm TBM}$ is associated with $m^{}_1$, only in the IO case can an $M^{}_{\rm D}$ of the form in Eq.~(\ref{3.1.5}) give rise to a TM1 mixing
\begin{eqnarray}
U = \left( \begin{array}{ccc} \vspace{0.15cm}
\displaystyle \frac{2}{\sqrt 6} & \displaystyle \frac{1}{\sqrt {3+ 2|r|^2 } } & \displaystyle \frac{2r^*}{\sqrt {6(3+ 2|r|^2)} }  \cr \vspace{0.15cm}
- \displaystyle \frac{1}{\sqrt 6} &  \displaystyle \frac{1 -r}{\sqrt {3+ 2|r|^2 } } & \displaystyle \frac{3+ 2 r^*}{\sqrt {6(3+ 2|r|^2)} }   \cr
\displaystyle \frac{1}{\sqrt 6} & - \displaystyle \frac{1 +r}{\sqrt {3+ 2|r|^2 } } & \displaystyle \frac{3- 2 r^*}{\sqrt {6(3+ 2|r|^2)} }  \cr
\end{array} \right) \left( \begin{matrix}
e^{{\rm i}\phi^{}_1} &  & \cr
& e^{{\rm i}\phi^{}_2 }  & \cr
&  & 1
\end{matrix} \right)  \;,
\label{3.1.6}
\end{eqnarray}
with $\phi^{}_1 = {\rm arg}(a)$ and $\phi^{}_2 = {\rm arg}(b)$.
In this case three light neutrino masses are given by
\begin{eqnarray}
m^{}_1 = 6|a|^2 \;, \hspace{1cm} m^{}_2 = |b|^2 (3+ 2|r|^2) \;, \hspace{1cm} m^{}_3 =0  \;.
\label{3.1.7}
\end{eqnarray}
The value of $r$ for $U$ in Eq.~(\ref{3.1.6}) to be compatible with the experimental results can be inferred from the following two relations
\begin{eqnarray}
s^{2}_{13} = \displaystyle \frac{ 2 |r|^2 }{3(3+ 2 |r|^2) } \;, \hspace{1cm}
\tan \theta^{}_{23} = \displaystyle \left| \frac{ 3 + 2 r }{ 3 - 2r } \right| \;.
\label{3.1.8}
\end{eqnarray}
For the $3\sigma$ ranges of $\theta^{}_{13}$ and $\theta^{}_{23}$,  $|r|$ and ${\rm arg}(r)$ are determined to be $0.31-0.34$ and $\pm(0.32-0.59)\pi$, respectively. This means that the second column of $M^{}_{\rm D}$ is dominantly contributed by $\langle \Phi^{}_2 \rangle$ rather than $\langle \Phi^{}_3 \rangle$.
Subsequently, the values of $\theta^{}_{12}$, $\delta$ and $\sigma-\rho$ can be calculated according to the formulas
\begin{eqnarray}
s^{2}_{12} =  \frac{ 1}{3} -  \frac{2 s^{2}_{13}}{3 - 3s^{2}_{13}} \;, \hspace{1cm}   \cos \delta = - \frac{1-5 s^2_{13}}{2 s^{}_{13} \tan{2\theta^{}_{23}} \sqrt{2 (1- 3 s^2_{13}) } } \;,
 \hspace{1cm} \sigma-\rho =  \phi^{}_2 -  \phi^{}_1  \;.
\label{3.1.9}
\end{eqnarray}
It is natural that, for the phases of $a$ and $b$, only their difference (i.e., $ \phi^{}_2 -  \phi^{}_1$) is of physical meaning.
At the $3\sigma$ level, $s^2_{12}$ and $\delta$ are respectively predicted to be $0.317-0.319$ and $\pm(0.33-0.59)\pi$, which are in good agreement with the experimental results. Finally, for the $3\sigma$ ranges of $\Delta m^2_{21}$ and $|\Delta m^2_{31}|$, $|a|^2$ and $|b|^2$ are determined to be $(8.11-8.42)\times 10^{-3}$ eV and $(1.53-1.60)\times 10^{-2}$ eV, respectively.

\subsubsection{TM2 scenario}

Similarly, one can achieve a TM2 mixing by having two columns of $M^{}_{\rm D}$ be respectively proportional and orthogonal to the second column of $U^{}_{\rm TBM}$ \cite{tminms}.
Such an $M^{}_{\rm D}$ can be parameterized as
\begin{eqnarray}
M^{}_{\rm D}= \left( \begin{array}{cc}
a \sqrt{M^{}_1} & 2 b r \sqrt{M^{}_2} \cr
a \sqrt{M^{}_1} &  b (1-r) \sqrt{M^{}_2} \cr
- a \sqrt{M^{}_1}  & b (1+r)  \sqrt{M^{}_2} \cr
\end{array} \right) \;,
\label{3.1.10}
\end{eqnarray}
which can be realized by associating $\langle \Phi^{}_2 \rangle$ with $N^{}_1$ and both $\langle \Phi^{}_1 \rangle$ and $\langle \Phi^{}_3 \rangle$ with $N^{}_2$. Now the dimensionless parameter $r$ measures the relative size between the contributions of $\langle \Phi^{}_1 \rangle$ and $\langle \Phi^{}_3 \rangle$ to the second column of $M^{}_{\rm D}$.
Since in the TM2 mixing the preserved second column of $U^{}_{\rm TBM}$ is associated with $m^{}_2$, a TM2 mixing can follow from the form of $M^{}_{\rm D}$ in Eq.~(\ref{3.1.10}) for both the NO and IO cases.

{\bf NO case:} In the NO case, the resulting TM2 mixing is given by
\begin{eqnarray}
U = \left( \begin{array}{ccc} \vspace{0.15cm}
\displaystyle \frac{2}{\sqrt {6(1+3|r|^2)} } & \displaystyle \frac{1}{\sqrt {3} } & \displaystyle \frac{2r}{\sqrt {2(1+ 3|r|^2)} }  \cr \vspace{0.15cm}
- \displaystyle \frac{1+3 r^*}{\sqrt {6(1+ 3 |r|^2)} }  &  \displaystyle \frac{1}{\sqrt {3 } } & \displaystyle \frac{1-r}{\sqrt {2(1+ 3|r|^2)} }   \cr
\displaystyle \frac{1-3 r^*}{\sqrt {6(1+ 3|r|^2)} } & - \displaystyle \frac{1}{\sqrt {3 } } & \displaystyle \frac{1+r}{\sqrt {2(1+ 3|r|^2)} } \cr
\end{array} \right) \left( \begin{matrix}
1 &  & \cr
& e^{{\rm i}\phi^{}_1 }  & \cr
&  & e^{{\rm i}\phi^{}_2}
\end{matrix} \right)  \;,
\label{3.1.11}
\end{eqnarray}
with also $\phi^{}_1 = {\rm arg}(a)$ and $\phi^{}_2 = {\rm arg}(b)$. And three light neutrino masses are given by
\begin{eqnarray}
m^{}_1 = 0 \;, \hspace{1cm} m^{}_2 = 3 |a|^2 \;, \hspace{1cm} m^{}_3 = 2 |b|^2 (1+ 3|r|^2)  \;.
\label{3.1.12}
\end{eqnarray}
The value of $r$ for $U$ in Eq.~(\ref{3.1.11}) to be compatible with the experimental results can be inferred from the following two relations
\begin{eqnarray}
s^{2}_{13} = \displaystyle \frac{ 2 |r|^2 }{1+ 3 |r|^2 } \;, \hspace{1cm}
\tan \theta^{}_{23} = \displaystyle \left| \frac{ 1 - r }{ 1+ r } \right| \;.
\label{3.1.13}
\end{eqnarray}
For the $3\sigma$ ranges of $\theta^{}_{13}$ and $\theta^{}_{23}$,  $|r|$ and ${\rm arg}(r)$ are respectively determined to be $0.10-0.11$ and $\pm(0.30-1.00)\pi$. This means that the second column of $M^{}_{\rm D}$ is dominantly contributed by $\langle \Phi^{}_3 \rangle$ rather than $\langle \Phi^{}_1 \rangle$.
Subsequently, the values of $\theta^{}_{12}$, $\delta$ and $\sigma$ can be calculated according to the formulas
\begin{eqnarray}
s^{2}_{12} = \frac{ 1}{3} + \frac{s^{2}_{13}}{3 - 3s^{2}_{13}} \;, \hspace{1cm} \cos \delta = \frac{1-2 s^2_{13}}{s^{}_{13} \tan{2\theta^{}_{23}} \sqrt{2- 3 s^2_{13}} } \;,
 \hspace{1cm} \sigma =  \phi^{}_1 - \phi^{}_2 - {\rm arg}(r) - \delta  \;.
\label{3.1.14}
\end{eqnarray}
At the $3\sigma$ level, $s^2_{12}$ and $\delta$ are respectively predicted to be $0.340-0.342$ and $\pm(0.307-1.00)\pi$. Finally, for the $3\sigma$ ranges of $\Delta m^2_{21}$ and $|\Delta m^2_{31}|$, $|a|^2$ and $|b|^2$ are determined to be $(2.78-3.01)\times 10^{-3}$ eV and $(2.39-2.50)\times 10^{-2}$ eV, respectively.

{\bf IO case:} For the IO case, the resulting TM2 mixing can be obtained from Eq.~(\ref{3.1.11}) by interchanging the first and third columns. And three light neutrino masses can be obtained from Eq.~(\ref{3.1.12}) by interchanging $m^{}_1$ and $m^{}_3$. Accordingly, the value of $r$ for the resulting neutrino mixing matrix to be compatible with the experimental results can be inferred from the following two relations
\begin{eqnarray}
s^{2}_{13} = \displaystyle \frac{2}{ 3(1+3|r|^2) } \;, \hspace{1cm}
\tan \theta^{}_{23} = \displaystyle \left| \frac{ 1 +3 r }{ 1 - 3 r } \right| \;.
\label{3.1.15}
\end{eqnarray}
For the $3\sigma$ ranges of $\theta^{}_{13}$ and $\theta^{}_{23}$,  $|r|$ and ${\rm arg}(r)$ are respectively determined to be $2.96-3.25$ and $\pm(0.00-0.68)\pi$. This means that the second column of $M^{}_{\rm D}$ is dominantly contributed by $\langle \Phi^{}_1 \rangle$ rather than $\langle \Phi^{}_3 \rangle$.
Subsequently, the values of $\theta^{}_{12}$ and $\delta$ can be calculated according to the same formulas as in Eq.~(\ref{3.1.14}) while the value of $\sigma -\rho$ can be obtained as $\sigma -\rho = \phi^{}_1 - \phi^{}_2 - {\rm arg}(r)$.
Finally, for the $3\sigma$ ranges of $\Delta m^2_{21}$ and $|\Delta m^2_{31}|$, $|a|^2$ and $|b|^2$ are determined to be $(1.65-1.71)\times 10^{-2}$ eV and $(7.45-9.25)\times 10^{-4}$ eV, respectively.

\subsection{RGE induced leptogenesis}

Now, in the scenario that two right-handed neutrinos have nearly degenerate low-scale masses, we consider the implications of the particular forms of $M^{}_{\rm D}$ in Eqs.~(\ref{3.1.5}, \ref{3.1.10}) for leptogenesis. It is immediate to see that, due to the orthogonality relation between their two columns (i.e., $(M^\dagger_{\rm D} M^{}_{\rm D})^{}_{IJ}  =0$ for $I \neq J$), one simply has $\varepsilon^{}_{I \alpha} =0$ (see Eq.~(\ref{2.1.2})), which preclude leptogenesis to proceed.

Fortunately, the renormalization group evolution effect is potentially capable of inducing a viable leptogenesis: for the forms of $M^{}_{\rm D}(\Lambda^{}_{\rm FS})$ in Eqs.~(\ref{3.1.5}, \ref{3.1.10}), its counterpart $M^{}_{\rm D}(M^{}_0)$ can be obtained as in Eq.~(\ref{2.2.2}). Despite its smallness (e.g., $\Delta^{}_{\tau} \simeq 1.65 \times 10^{-5}$ for $M^{}_0 \simeq 1$ TeV and $\Lambda^{}_{\rm FS} \simeq 10^{10}$ GeV), $\Delta^{}_\tau$ can bring about a dramatic effect for leptogenesis: the orthogonality relation between two columns of $M^{}_{\rm D}(\Lambda^{}_{\rm FS})$ will be eventually broken, making leptogenesis possible \cite{king, xingzhang}. On the other hand,
due to $(Y^\dagger_\nu Y^{}_\nu)^{}_{12} =0$, the off-diagonal terms of $M^{}_{\rm R} (M^{}_0)$ in Eq.~(\ref{2.2.4}) are vanishing.

\subsubsection{TM1 scenario}

For an $M^{}_{\rm D}(\Lambda^{}_{\rm FS})$ of the form in Eq.~(\ref{3.1.5}) which gives a TM1 mixing in the IO case, the renormalization group evolution effect will induce non-vanishing $\varepsilon^{}_{\alpha}$ as
\begin{eqnarray}
&& \varepsilon^{}_{e} \simeq  \Delta^{}_\tau \frac{ 128 \pi v^2 m^{}_0 M^2_0 \Delta M  }{256 \pi^2 v^4 \Delta M^2 + m^{2}_0 M^4_0} \frac{ {\rm Re}[(1+r) e^{{\rm i} \phi^{}_{21}}] }{3 (3+2 |r|^2)} {\rm Im}( e^{{\rm i} \phi^{}_{21} }) \;, \nonumber \\
&& \varepsilon^{}_{\mu} \simeq - \Delta^{}_\tau \frac{ 64 \pi v^2 m^{}_0 M^2_0 \Delta M  }{256 \pi^2 v^4 \Delta M^2 + m^{2}_0 M^4_0} \frac{ {\rm Re}[(1+r) e^{{\rm i} \phi^{}_{21} } ] }{3(3+2|r|^2)} {\rm Im}[(1-r) e^{{\rm i} \phi^{}_{21} }] \;, \nonumber \\
&& \varepsilon^{}_{\tau} \simeq - \Delta^{}_\tau \frac{ 64 \pi v^2 m^{}_0 M^2_0 \Delta M   }{256 \pi^2 v^4 \Delta M^2 + m^{2}_0 M^4_0} \frac{ {\rm Re}[ (1+r) e^{{\rm i} \phi^{}_{21} } ] }{3(3+2|r|^2) } {\rm Im}[(1+r) e^{{\rm i} \phi^{}_{21} }] \;,
\label{3.2.1}
\end{eqnarray}
where $M^{}_0 \equiv (M^{}_1 + M^{}_2)/2$, $m^{}_0 \equiv (m^{}_1 + m^{}_2)/2$ and $\phi^{}_{21} \equiv \phi^{}_2 - \phi^{}_1$ have been defined. In obtaining these results, we have made use of the near degeneracy between $M^{}_1$ and $M^{}_2$ (i.e., $M^{}_1 \simeq M^{}_2 \simeq M^{}_0$) and $m^{}_1$ and $m^{}_2$ (i.e., $m^{}_1 \simeq m^{}_2 \simeq m^{}_0$) to simplify the expressions. One can see that the magnitudes of $\varepsilon^{}_{\alpha}$ are directly controlled by $\Delta^{}_\tau$. And the relation $\varepsilon^{}_{e} + \varepsilon^{}_{\mu} + \varepsilon^{}_{\tau} \simeq 0$ holds. This is because the total CP asymmetry
\begin{eqnarray}
\varepsilon^{}_I = \varepsilon^{}_{I e} + \varepsilon^{}_{I \mu} + \varepsilon^{}_{I \tau} = \frac{{\rm Im}\left[(M^\dagger_{\rm D}
M^{}_{\rm D})^2_{IJ}\right]}{8\pi  v^2 (M^\dagger_{\rm D} M^{}_{\rm D})^{}_{II} } \cdot \frac{M^{}_I M^{}_J \Delta M^2_{IJ}}{(\Delta M^2_{IJ})^2 + M^2_I \Gamma^2_J} \; ,
\label{3.2.2}
\end{eqnarray}
is proportional to ${\rm Im}\left[(M^\dagger_{\rm D} M^{}_{\rm D})^2_{IJ} \right]$ which only becomes non-vanishing at the order of $\Delta^2_\tau$. Furthermore, it is useful to note that $\Delta M$ and $M^{}_0$ only take effect in the form of $\Delta M/M^2_0$ in $\varepsilon^{}_\alpha$. And the dependence of $\varepsilon^{}_\alpha$ on $\phi^{}_{21}$ has a period of $\pi$ (e.g., $\varepsilon^{}_e \propto [1+{\rm Re}(r)] \cos \phi^{}_{21} \sin \phi^{}_{21} - {\rm Im}(r) \sin^2 \phi^{}_{21}$). Taking account of the smallness of $r$ (as obtained below Eq.~(\ref{3.1.8})), one arrives at $\varepsilon^{}_{e} \sim - 2 \varepsilon^{}_{\mu} \sim -2 \varepsilon^{}_{\tau}$.
On the other hand, the washout mass parameters $\widetilde m^{}_\alpha$ are obtained as
\begin{eqnarray}
&& \widetilde m^{}_e = \frac{2 }{3} m^{}_1 + \frac{ 1}{3+2 |r|^2} m^{}_2 \simeq (0.048-0.050) \ {\rm eV} \;, \nonumber \\
&& \widetilde m^{}_\mu = \frac{ 1 }{6} m^{}_1 + \frac{ |1-r|^2 }{3+2 |r|^2} m^{}_2 \simeq (0.020-0.029) \ {\rm eV} \;, \nonumber \\
&& \widetilde m^{}_\tau = \frac{ 1 }{6} m^{}_1 + \frac{ |1+r|^2 }{3+2 |r|^2} m^{}_2 \simeq (0.023-0.032) \ {\rm eV} \;.
\label{3.2.3}
\end{eqnarray}

With these results, one can now calculate the final baryon asymmetry. In order to show the dependence of $Y^{}_{\rm B}$ on $\phi^{}_{21}$ and $D  \equiv \Delta M/M^{}_0$, in Fig.~1(a) and (b) (for $D > 0$ and $D < 0$, respectively) we have plotted the contour lines of $Y^{}_{\rm B}/Y^0_{\rm B}$ on the $\phi^{}_{21}$-$D$ plane. Here we have only shown the contour lines for $Y^{}_{\rm B} \ge 0$. Note that $Y^{}_{\rm B}$ would undergo a sign reversal under the transformation $D \to -D$ (see Eq.~(\ref{3.2.1})).
In obtaining these results (same as below, unless otherwise specified) we have taken $M^{}_0= 1$ TeV, $\Lambda^{}_{\rm FS} = 10^{10}$ GeV and the best-fit values of the neutrino oscillation parameters as typical inputs. On the one hand, the results of $Y^{}_{\rm B}$ for other values of $M^{}_0$ can be inferred with the help of the aforementioned observation that $D$ and $M^{}_0$ only take effect in the form of $D/M^{}_0$ (i.e., $\Delta M/M^2_0$) in $\varepsilon^{}_\alpha$. On the other hand, the results of $Y^{}_{\rm B}$ for other values of $\Lambda^{}_{\rm FS}$ can be inferred from the near logarithmic dependence of $\Delta^{}_\tau$ on $\Lambda^{}_{\rm FS}$ as in Eq.~(\ref{2.2.3}). From these results one can see that $Y^{}_{\rm B}$ has no chance to reach $Y^0_{\rm B}$. After taking account of the $3\sigma$ ranges of the neutrino oscillation parameters, it is found that the maximally allowed value $Y^{\rm max}_{\rm B}$ of $Y^{}_{\rm B}$ can only reach 58\% of $Y^0_{\rm B}$. And $Y^{\rm max}_{\rm B}$ is achieved at $(\phi^{}_{21}, D) \simeq (0.25\pi, 1.6 \times 10^{-14})$, which can be easily understood from Eq.~(\ref{3.2.1}): the $\Delta M$-dependent part will take its maximally allowed value when $16 \pi v^2 \Delta M = m^{}_0 M^2_0$ (i.e., $D = m^{}_0 M^{}_0/(16 \pi v^2)$) holds, while the $\phi^{}_{21}$-dependent part is approximately proportional to $\cos \phi^{}_{21} \sin \phi^{}_{21}$ (as a result of the smallness of $r$). Even if $\Lambda^{}_{\rm FS}$ happens to be close to the grand unification scale $10^{15}$ GeV in which case $\Delta^{}_\tau$ will get enhanced by about 70\%, it is still almost impossible for $Y^{}_{\rm B}$ to reach $Y^0_{\rm B}$. But in the MSSM framework where $\Delta^{}_\tau$ will get enhanced by about the factor $-2(1+\tan^2 \beta)/3$, to make a successful leptogenesis possible, one just needs to have $\tan \beta \geq 1.3$ (and similarly for the following discussions).

Finally, we consider the case that two right-handed neutrinos are exactly degenerate at the flavor-symmetry scale (i.e., $\Delta M^0_{21} =0$). In this case, their mass difference is completely due to the renormalization group evolution effect:
\begin{eqnarray}
\Delta M \simeq  - \frac{1}{8\pi^2 v^2} (m^{}_2 - m^{}_1) M^2_0 \ln\left( \frac{\Lambda^{}_{\rm FS}}{M^{}_0} \right) \simeq - (2.3-2.8) \times 10^{-3} \ {\rm eV} \;,
\label{3.2.4}
\end{eqnarray}
corresponding to $D\simeq - (2.3-2.8) \times 10^{-15}$. For this value of $\Delta M$, the maximally allowed value $Y^{\rm max}_{\rm B}$ of $Y^{}_{\rm B}$ can only reach 9.3\% of $Y^0_{\rm B}$.

%%%%%%%%%%%%%%%%%%%%%% Figure 4-1 %%%%%%%%%%%%%%%%%%%%%%
\begin{figure*}[t]
\centering
\includegraphics[width=6in]{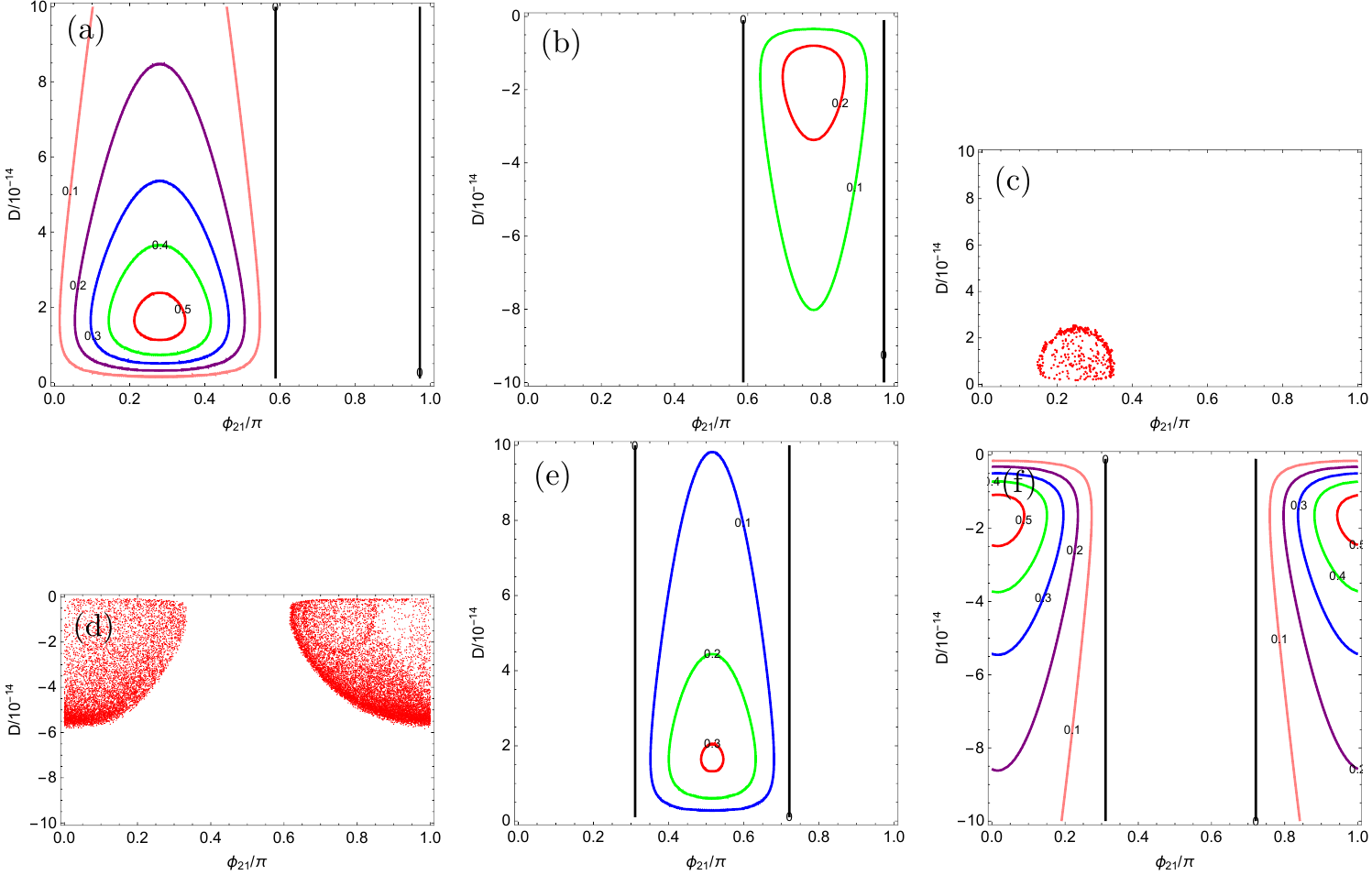}
\caption{ The contour lines of $Y^{}_{\rm B}/Y^{0}_{\rm B}$ on the $\phi^{}_{21}$-$D$ plane: the IO case with the TM1 mixing for $D>0$ (a) and $D<0$ (b); the IO case with the TM2 mixing for $D>0$ (e) and $D<0$ (f). The values of $\phi^{}_{21}$ and $D$ that allow $Y^{}_{\rm B}$ to reproduce $Y^{0}_{\rm B}$: the NO case with the TM2 mixing for $D>0$ (c) and $D<0$ (f). }
\label{Fig:4-1}
\end{figure*}
%%%%%%%%%%%%%%%%%%%%%%%%%%%%%%%%%%%%%%%%%%%%%%%%%%%

\subsubsection{TM2 scenario}

{\bf NO case:} For the TM2 mixing scenario, similar results are obtained.
For an $M^{}_{\rm D}(\Lambda^{}_{\rm FS})$ of the form in Eq.~(\ref{3.1.10}) which yields a TM2 mixing, in the NO case the renormalization group evolution effect will induce non-vanishing $\varepsilon^{}_{\alpha}$ as
\begin{eqnarray}
&& \hspace{ -1cm}  \varepsilon^{}_{e}  \simeq \Delta^{}_\tau  \left[ \frac{64 \pi v^2 m^{}_2 M^2_0 \Delta M  }{ 256 \pi^2 v^4 \Delta M^2+ m^2_2 M^4_0  } + \frac{64 \pi v^2 m^{}_3 M^2_0 \Delta M  }{ 256 \pi^2 v^4 \Delta M^2+ m^2_3 M^4_0  }  \right] \frac{ {\rm Re}[(1+r) e^{{\rm i} \phi^{}_{21} } ] }{3(1+3|r|^2)} {\rm Im}(r e^{{\rm i} \phi^{}_{21} })  \;, \nonumber \\
&& \hspace{ -1cm} \varepsilon^{}_{\mu} \simeq  \Delta^{}_\tau   \left[ \frac{32 \pi v^2 m^{}_2 M^2_0 \Delta M}{256 \pi^2 v^4 \Delta M^2+ m^2_2 M^4_0 } +  \frac{32 \pi v^2 m^{}_3 M^2_0 \Delta M}{256 \pi^2 v^4 \Delta M^2+ m^2_3 M^4_0 } \right] \frac{{\rm Re}[(1+r)e^{{\rm i} \phi^{}_{21} } ] }{3(1+3|r|^2)} {\rm Im}[(1-r)e^{{\rm i} \phi^{}_{21} } ] \;, \nonumber \\
&& \hspace{ -1cm} \varepsilon^{}_{\tau} \simeq - \Delta^{}_\tau   \left[ \frac{32 \pi v^2 m^{}_2 M^2_0 \Delta M}{256 \pi^2 v^4 \Delta M^2+ m^2_2 M^4_0 } +  \frac{32 \pi v^2 m^{}_3 M^2_0 \Delta M}{256 \pi^2 v^4 \Delta M^2+ m^2_3 M^4_0 } \right] \frac{{\rm Re}[(1+r)e^{{\rm i} \phi^{}_{21} } ] }{3(1+3|r|^2)} {\rm Im}[(1+ r)e^{{\rm i} \phi^{}_{21} } ] \;.
\label{3.2.5}
\end{eqnarray}
It is easy to see that the above observations for the results for the TM1 mixing scenario still hold.
Due to the smallness of $r$ (as obtained below Eq.~(\ref{3.1.13})), one generally has $\varepsilon^{}_\mu \simeq - \varepsilon^{}_\tau$ and $|\varepsilon^{}_\mu| \gg |\varepsilon^{}_e|$ except for the case of $\phi^{}_{21} \simeq 0$ or $\pi$.
On the other hand, the washout mass parameters $\widetilde m^{}_\alpha$ are given by
\begin{eqnarray}
&& \widetilde m^{}_e = \frac{1}{3} m^{}_2 + \frac{ 2 |r|^2 }{ 1+ 3|r|^2 } m^{}_3 \simeq (0.0038-0.0042) \ {\rm eV} \;, \nonumber \\
&& \widetilde m^{}_\mu = \frac{1}{3} m^{}_2 + \frac{ |1-r|^2 }{ 2(1+3 |r|^2)} m^{}_3 \simeq (0.024-0.034) \ {\rm eV} \;, \nonumber \\
&& \widetilde m^{}_\tau = \frac{1}{3} m^{}_2 + \frac{ |1+r|^2 }{2(1+3 |r|^2) } m^{}_3 \simeq (0.022-0.031) \ {\rm eV} \;.
\label{3.2.6}
\end{eqnarray}
For the following two reasons, $Y^{}_{\rm B}$ is dominantly contributed by $ \varepsilon^{}_{e}  \kappa ( \widetilde m^{}_e)$ rather than $ \varepsilon^{}_{\mu}  \kappa ( \widetilde m^{}_\mu ) + \varepsilon^{}_{\tau}  \kappa ( \widetilde m^{}_\tau )$. On the one hand, as a result of the near equality between $\widetilde m^{}_\mu$ and $\widetilde m^{}_\tau$, a heavy cancellation between $\varepsilon^{}_{\mu}  \kappa ( \widetilde m^{}_\mu )$ and  $\varepsilon^{}_{\tau}  \kappa ( \widetilde m^{}_\tau )$ happens: $ \varepsilon^{}_{\mu}  \kappa ( \widetilde m^{}_\mu ) + \varepsilon^{}_{\tau}  \kappa ( \widetilde m^{}_\tau ) \simeq (\varepsilon^{}_{\mu} +  \varepsilon^{}_{\tau} )  \kappa ( \widetilde m^{}_\mu) \simeq - \varepsilon^{}_{e}  \kappa ( \widetilde m^{}_\mu )$. On the other hand, since $\kappa ( \widetilde m^{}_\alpha )$ is roughly inversely proportional to $\widetilde m^{}_\alpha$ for $\widetilde m^{}_\alpha \gtrsim 3 m^{}_*$ (with $ m^{}_* \simeq 1.1 \times 10^{-3}$ eV) \cite{Lreview}, which can be seen from the following analytical fit of $\kappa \left( \widetilde m^{}_\alpha \right)$ \cite{giudice}
\begin{eqnarray}
\frac{1}{\kappa( \widetilde m^{}_\alpha )} \simeq \frac{3.3 \times 10^{-3} ~{\rm eV}}{ \widetilde m^{}_\alpha} + \left( \frac{ \widetilde m^{}_\alpha } {5.5 \times 10^{-4} ~{\rm eV}} \right)^{1.16} \;,
\label{3.2.7}
\end{eqnarray}
$\kappa ( \widetilde m^{}_e)$ is much larger than $\kappa ( \widetilde m^{}_\mu )$ as a result of $\widetilde m^{}_\mu \gg \widetilde m^{}_e \gtrsim 3 m^{}_*$. Numerically, it is found that $Y^{}_{\rm B}$ has chance to reproduce $Y^0_{\rm B}$. Figure~1(c) and (d) (for $D>0$ and $D<0$, respectively) show the values of $\phi^{}_{21}$ and $D$ for realizing $Y^{}_{\rm B}= Y^0_{\rm B}$ within the $3\sigma$ level.

In the case that two right-handed neutrinos are exactly degenerate at the flavor-symmetry scale, their mass difference induced by the renormalization group evolution effect is given by
\begin{eqnarray}
\Delta M \simeq  \frac{1}{8\pi^2 v^2} (m^{}_2 - m^{}_3) M^2_0 \ln\left( \frac{\Lambda^{}_{\rm FS}}{M^{}_0} \right) \simeq - (0.14-0.16) \ {\rm eV} \;,
\label{3.2.8}
\end{eqnarray}
corresponding to $D \simeq - (1.4-1.6) \times 10^{-13}$. For this value of $\Delta M$, the maximally allowed value $Y^{\rm max}_{\rm B}$ of $Y^{}_{\rm B}$ can only reach 43\% of $Y^0_{\rm B}$. To have a successful leptogenesis, $\Delta M^0_{21}$ needs to be non-vanishing so that it can partly cancel the contribution of the renormalization group evolution effect to $\Delta M$ and finally give $\Delta M \geq - 0.057$ eV (see Fig.~1(d)).

{\bf IO case:} In the IO case, the corresponding results can be obtained from those in the NO case by making the replacements $m^{}_2 \to m^{}_1$ and $m^{}_3 \to m^{}_2$. To be explicit,
the non-vanishing $\varepsilon^{}_{\alpha}$ induced by the renormalization group evolution effect become
\begin{eqnarray}
&& \varepsilon^{}_{e} \simeq  \Delta^{}_\tau \frac{128 \pi v^2 m^{}_0 M^2_0 \Delta M }{256 \pi^2 v^4 \Delta M^2+ m^2_0 M^4_0} \frac{ {\rm Re}[(1+r) e^{{\rm i} \phi^{}_{21} }  ] }{ 3 (1+3 |r|^2) }  {\rm Im}( re^{{\rm i} \phi^{}_{21}  } )  \;, \nonumber \\
&& \varepsilon^{}_{\mu} \simeq  \Delta^{}_\tau \frac{64 \pi v^2 m^{}_0 M^2_0 \Delta M }{256 \pi^2 v^4 \Delta M^2+ m^2_0 M^4_0} \frac{ {\rm Re}[(1+r)  e^{{\rm i} \phi^{}_{21}  } ] }{3(1+3|r|^2)} {\rm Im}[(1-r) e^{{\rm i} \phi^{}_{21}  }  ]  \;, \nonumber \\
&& \varepsilon^{}_{\tau} \simeq  - \Delta^{}_\tau \frac{64 \pi v^2 m^{}_0 M^2_0 \Delta M }{256 \pi^2 v^4 \Delta M^2+ m^2_0 M^4_0} \frac{ {\rm Re}[(1+r)  e^{{\rm i} \phi^{}_{21}  } ] }{3(1+3|r|^2)} {\rm Im}[(1+ r) e^{{\rm i} \phi^{}_{21} }  ]  \;.
\label{3.2.9}
\end{eqnarray}
Taking account of the largeness of $r$ (as obtained below Eq.~(\ref{3.1.15})), one can see that these results are nearly same as those in Eq.~(\ref{3.2.1}) except for the replacement $\phi^{}_{21} \to {\rm arg}(r) + \phi^{}_{21}$.
On the other hand, the washout mass parameters $\widetilde m^{}_\alpha$ turn out to be
\begin{eqnarray}
&& \widetilde m^{}_e = \frac{ 2 |r|^2 }{ 1+ 3|r|^2 } m^{}_1 + \frac{1}{3} m^{}_2  \simeq (0.048-0.050) \ {\rm eV}  \;, \nonumber \\
&& \widetilde m^{}_\mu =  \frac{ |1-r|^2 }{ 2(1+3 |r|^2)} m^{}_1 + \frac{1}{3} m^{}_2 \simeq (0.020-0.029) \ {\rm eV}  \;, \nonumber \\
&& \widetilde m^{}_\tau =  \frac{ |1+r|^2 }{2(1+3 |r|^2) } m^{}_1 + \frac{1}{3} m^{}_2 \simeq (0.023-0.031) \ {\rm eV}  \;.
\label{3.2.10}
\end{eqnarray}
Consequently, the results of $Y^{}_{\rm B}$ are nearly same as those in the TM1 mixing scenario (see Fig.~1(e) and (f)). It is found that the maximally allowed value $Y^{\rm max}_{\rm B}$ of $Y^{}_{\rm B}$ can only reach 55\% of $Y^{0}_{\rm B}$, which is achieved at $(\phi^{}_{21}, D) \simeq (0, -1.6 \times 10^{-14})$.

In the case that two right-handed neutrinos are exactly degenerate at the flavor-symmetry scale, their mass difference induced by the renormalization group evolution effect takes a value opposite to that in Eq.~(\ref{3.2.4}). For such a value of $\Delta M$, the maximally allowed value $Y^{\rm max}_{\rm B}$ of $Y^{}_{\rm B}$ can only reach 12\% of $Y^0_{\rm B}$.

\section{$\mu$-$\tau$ reflection symmetry scenario}

\subsection{$\mu$-$\tau$ reflection symmetry in minimal seesaw}

To implement the $\mu$-$\tau$ reflection symmetry in the minimal seesaw model, we specify the transformation properties of two right-handed neutrino fields under it as
$N^{}_1 \leftrightarrow N^{c}_1$ and $N^{}_2 \leftrightarrow N^{c}_2$ \cite{mutaums} \footnote{See Ref.~\cite{mutaums2} for an alternative choice: $N^{}_1 \leftrightarrow N^{c}_2$ and $N^{}_2 \leftrightarrow N^{c}_1$ which can be viewed as a generalization of the $\mu$-$\tau$ reflection symmetry from the left-handed neutrino sector to the right-handed neutrino sector.}. In this case, the symmetry requirement (that the neutrino mass matrix should keep invariant under the above transformations of two right-handed neutrino fields and those of three left-handed neutrino fields in Eq.~(\ref{4})) restricts the Dirac neutrino mass matrix to a form as
\begin{eqnarray}
M^{(0)}_{\rm D} = \left( \begin{array}{cc}
c  & d  \cr
e  & f   \cr
e^* & f^*  \cr
\end{array} \right)  \;,
\label{4.1.1}
\end{eqnarray}
where $c$ and $d$ are real parameters while $e$ and $f$ are generally complex parameters. On the other hand, the diagonal right-handed neutrino mass matrix is restricted to a form as $D^{(0)}_{\rm R} = {\rm diag}(\pm M^{}_1, \pm M^{}_2)$. After a transformation to the basis of $D^{}_{\rm R} = {\rm diag}(M^{}_1, M^{}_2)$ by means of the diagonal phase matrix $P^{}_N = {\rm diag}(\eta^{}_1, \eta^{}_2)$ (for $\eta^{}_I = 1$ or i, corresponding to the ``$+$" or ``$-$" sign of $M^{}_I$ in $D^{(0)}_{\rm R}$), the Dirac neutrino mass matrix has a final form as $M^{}_{\rm D} = M^{(0)}_{\rm D} P^{}_N$.

Then, we consider the question of how to connect the parameters $c$, $d$, $e$ and $f$ with the low-energy neutrino parameters.
As is known, in the Casas-Ibarra parametrization \cite{CI}, the Dirac neutrino mass matrix can be generally expressed as
\begin{eqnarray}
M^{}_{\rm D} = {\rm i} U \sqrt{D^{}_\nu} R \sqrt{D^{}_{\rm R}}  \;,
\label{4.1.2}
\end{eqnarray}
where $R$ is a $3\times 2$ matrix as
\begin{eqnarray}
{\rm NO}: \hspace{1cm} R = \left( \begin{array}{cc} 0 & 0 \cr \cos{z} & \sin{z} \cr - \sin{z} & \cos{z} \end{array} \right) \;;  \hspace{1cm}
{\rm IO}: \hspace{1cm} R = \left( \begin{array}{cc} \cos{z} & \sin{z} \cr - \sin{z} &  \cos{z} \cr 0 & 0 \end{array} \right) \;,
\label{4.1.3}
\end{eqnarray}
with $z$ being a complex parameter.
To be explicit, the elements of $M^{}_{\rm D}$ read
\begin{eqnarray}
&& \left(M^{}_{\rm D}\right)^{}_{\alpha 1}  = {\rm i} \sqrt{M^{}_1} \left( U^{}_{\alpha i} \sqrt{m^{}_i} \cos z -  U^{}_{\alpha j} \sqrt{m^{}_j} \sin z \right) \;, \nonumber \\
&& \left(M^{}_{\rm D}\right)^{}_{\alpha 2} =   {\rm i} \sqrt{M^{}_2}  \left( U^{}_{\alpha i} \sqrt{m^{}_i} \sin z +  U^{}_{\alpha j} \sqrt{m^{}_j} \cos z \right) \;,
\label{4.1.4}
\end{eqnarray}
with $i=2$ and $j=3$ ($i=1$ and $j=2$) in the NO (IO) case.
In the presence of the $\mu$-$\tau$ reflection symmetry, the special form of $M^{}_{\rm D} = M^{(0)}_{\rm D} P^{}_N$ yields some constraints on the parameters of $U$ and $z$. While the constraints on the parameters of $U$ have already been given in Eq.~(\ref{5}), the constraint on $z$ can be derived via a direct comparison between ${\rm i} U \sqrt{D^{}_\nu} R \sqrt{D^{}_{\rm R}}$ with $M^{(0)}_{\rm D} P^{}_N$. A careful analysis reveals that there are the following possible cases: (1) in the NO case with $\sigma =\pi/2$ and $P^{}_N = {\rm diag}({\rm i}, {\rm i})$, one arrives at $\cos z = \cos \theta$ and $\sin z = \sin \theta$ with $\theta$ being a real parameter. (2) In the IO case with $\sigma -\rho =0$ and $P^{}_N = {\rm diag}(1, 1)$, one also has $\cos z = \cos \theta$ and $\sin z = \sin \theta$.  (3) In the NO case with $\sigma =0$ and $P^{}_N = {\rm diag}(1, {\rm i})$ (or ${\rm diag}({\rm i}, 1)$), the results become $\cos z = \cosh \theta$ and $\sin z = {\rm i} \sinh \theta $ (or $\cos z = {\rm i} \sinh \theta$ and $\sin z = \cosh \theta$). (4) In the IO case with $\sigma-\rho = \pi/2$ and $P^{}_N = {\rm diag}(1, {\rm i})$ (or ${\rm diag}({\rm i}, 1)$), one also has $\cos z = \cosh \theta$ and $\sin z = {\rm i} \sinh \theta $ (or $\cos z = {\rm i} \sinh \theta$ and $\sin z = \cosh \theta$).

\subsection{RGE induced leptogenesis}

Now, still in the scenario that two right-handed neutrinos have nearly degenerate TeV-scale masses, let us consider the implication of the $\mu$-$\tau$ reflection symmetry (i.e., the special form of $M^{}_{\rm D} = M^{(0)}_{\rm D} P^{}_N$) for leptogenesis. In the case of $P^{}_N ={\rm diag} (1, 1)$ or ${\rm diag}({\rm i}, {\rm i})$, it is direct to see that $\varepsilon^{}_{e} =0$ and $\varepsilon^{}_{\tau} = - \varepsilon^{}_{\mu}$.
In this case the final baryon asymmetry is given by $Y^{}_{\rm B} = c r \varepsilon^{}_{\mu} [\kappa ( \widetilde m^{}_\mu) - \kappa ( \widetilde m^{}_\tau ) ]$. Because of $\widetilde m^{}_\mu = \widetilde m^{}_\tau$, $Y^{}_{\rm B}$ will vanish completely, rendering leptogenesis unworkable. In the case of $P^{}_N = {\rm diag} (1, {\rm i})$ or ${\rm diag} ({\rm i}, 1)$, one simply has $\varepsilon^{}_{\alpha} =0$, which can be understood from
\begin{eqnarray}
&& \varepsilon^{}_{I\alpha} \propto {\rm Im}\left\{ (M^*_{\rm D})^{}_{\alpha I} (M^{}_{\rm D})^{}_{\alpha J} M^{}_0 \left[ (M^\dagger_{\rm D} M^{}_{\rm D})^{}_{IJ} + (M^\dagger_{\rm D} M^{}_{\rm D})^{}_{JI} \right] \right\}  \nonumber \\
&& \hspace{0.65cm} =  {\rm Im}\left\{ (M^*_{\rm D})^{}_{\alpha I} (M^{}_{\rm D})^{}_{\alpha J} M^{}_0 \left[ - {\rm i} (c d + e^* f+ e f^*) + {\rm i} (c d + e f^* + e^* f) \right] \right\} =0 \;,
\label{4.2.1}
\end{eqnarray}
so leptogenesis cannot work either.

Like in the trimaximal mixing scenario, the renormalization group evolution effect can also play a dramatic role for leptogenesis in the present scenario: given an $M^{}_{\rm D}(\Lambda^{}_{\rm FS})$ of the $\mu$-$\tau$ reflection symmetry at the flavor-symmetry scale, its counterpart $M^{}_{\rm D}(M^{}_0)$ at the right-handed neutrino mass scale can be obtained as in Eq.~(\ref{2.2.2}) \cite{zhou}. Thanks to the difference between $\Delta^{}_\mu$ and $\Delta^{}_\tau$, the $\mu$-$\tau$ reflection symmetry will be broken, thus allowing leptogenesis to proceed.

\subsubsection{$P^{}_N ={\rm diag} (1, 1)$ or ${\rm diag}({\rm i}, {\rm i})$}

In the case of $P^{}_N ={\rm diag} (1, 1)$ or ${\rm diag}({\rm i}, {\rm i})$, the right-handed neutrino mass matrix at the right-handed neutrino mass scale given by Eq.~(\ref{2.2.4}) turns out to be
\begin{eqnarray}
M^{}_{\rm R} (M^{}_0) \simeq
\left( \begin{array}{cc}
M^{0}_1 - \displaystyle \frac{M^{}_0}{8\pi^2 v^2} ( c^2 + 2 |e|^2) \ln\left( \frac{\Lambda^{}_{\rm FS}}{M^{}_0} \right)  &  - \displaystyle \frac{M^{}_0}{8\pi^2 v^2} [ c d + 2{\rm Re}(e^* f) ] \ln\left( \frac{\Lambda^{}_{\rm FS}}{M^{2}_0} \right)   \cr
- \displaystyle \frac{M^{}_0}{8\pi^2 v^2} [ c d + 2{\rm Re}(e^* f) ] \ln\left( \frac{\Lambda^{}_{\rm FS}}{M^{}_0} \right) & M^{0}_2 - \displaystyle \frac{M^{}_0}{8\pi^2 v^2} ( d^2 + 2 |f|^2) \ln\left( \frac{\Lambda^{}_{\rm FS}}{M^{}_0} \right)  \cr
\end{array} \right) \;.
\label{4.2.2}
\end{eqnarray}
One can go back to the mass basis of the right-handed neutrinos via a unitary (actually orthogonal) transformation as
\begin{eqnarray}
U^{}_{\rm R} =
\left( \begin{array}{cc}
\cos \theta^{}_{\rm R}  &  \sin \theta^{}_{\rm R}  \cr
- \sin \theta^{}_{\rm R} & \cos \theta^{}_{\rm R} \cr
\end{array} \right) \;,
\label{4.2.3}
\end{eqnarray}
with
\begin{eqnarray}
\tan 2 \theta^{}_{\rm R} = \frac{ -2 M^{}_0 [ c d + 2{\rm Re}(e^* f) ] \ln\left( \displaystyle  \frac{\Lambda^{}_{\rm FS}}{M^{}_0} \right) }{ 8 \pi^2 v^2 \Delta M^0_{21} + M^{}_0 (c^2 + 2 |e|^2 - d^2 - 2 |f|^2 ) \ln\left( \displaystyle  \frac{\Lambda^{}_{\rm FS}}{M^{}_0} \right) } \;.
\label{4.2.4}
\end{eqnarray}
And the right-handed neutrino mass difference is given by
\begin{eqnarray}
& & \Delta M \simeq  \left[ \Delta M^0_{21} + \frac{ M^{}_0} {8 \pi^2 v^2 } (c^2 + 2 |e|^2 - d^2 - 2 |f|^2 ) \ln\left( \displaystyle  \frac{\Lambda^{}_{\rm FS}}{M^{}_0} \right) \right] \cos 2 \theta^{}_{\rm R} \nonumber \\
& & \hspace{1.3cm} - \displaystyle \frac{M^{}_0}{4\pi^2 v^2} [ c d + 2{\rm Re}(e^* f) ] \ln\left( \frac{\Lambda^{}_{\rm FS}}{M^{}_0} \right) \sin 2 \theta^{}_{\rm R} \;.
\label{4.2.5}
\end{eqnarray}

On the other hand, the Dirac neutrino mass matrix at the right-handed neutrino mass scale is transformed to a form as
\begin{eqnarray}
M^{\prime}_{\rm D}(M^{}_0) = M^{}_{\rm D} (M^{}_0) U^{}_{\rm R}
\simeq I^{}_0 \left( \begin{array}{ccc}
1 &   &  \cr
 & 1  &  \cr
 &  & 1+\Delta^{}_{\tau} \cr
\end{array} \right)
\left( \begin{array}{cc}
c  & d  \cr
e  & f   \cr
e^* & f^*  \cr
\end{array} \right) \left( \begin{array}{cc}
\cos \theta^{}_{\rm R}  &  \sin \theta^{}_{\rm R}  \cr
- \sin \theta^{}_{\rm R} & \cos \theta^{}_{\rm R} \cr
\end{array} \right) P^{}_N \;.
\label{4.2.6}
\end{eqnarray}
If we introduce the following parameters
\begin{eqnarray}
c^\prime = c \cos \theta^{}_{\rm R} - d \sin \theta^{}_{\rm R} \;, \hspace{1cm}
d^\prime = c \sin \theta^{}_{\rm R} + d \cos \theta^{}_{\rm R} \;, \nonumber \\
e^\prime = e \cos \theta^{}_{\rm R} - f \sin \theta^{}_{\rm R} \;, \hspace{1cm}
f^\prime = e \sin \theta^{}_{\rm R} + f \cos \theta^{}_{\rm R} \;,
\label{4.2.7}
\end{eqnarray}
then $M^{\prime}_{\rm D}(M^{}_0)$ will take a same form as $M^{}_{\rm D}(\Lambda^{}_{\rm FS})$ (except for the small factor $\Delta^{}_\tau$):
\begin{eqnarray}
M^{\prime}_{\rm D}(M^{}_0) \simeq I^{}_0
\left( \begin{array}{ccc}
1 &   &  \cr
 & 1  &  \cr
 &  & 1+\Delta^{}_{\tau} \cr
\end{array} \right)
\left( \begin{array}{cc}
c^\prime  & d^\prime  \cr
e^\prime  & f^\prime   \cr
e^{\prime *} & f^{\prime *}  \cr
\end{array} \right)  P^{}_N \;.
\label{4.2.8}
\end{eqnarray} 
By inserting such an $M^{\prime}_{\rm D}(M^{}_0)$ into Eq.~(\ref{2.1.2}), one obtains $\varepsilon^{}_{e} =0$,
\begin{eqnarray}
&& \varepsilon^{}_{\mu} \simeq   -[ c^\prime d^\prime + 2{\rm Re}(e^{\prime *} f^\prime) ] {\rm Im}(e^{\prime *} f^\prime) \nonumber \\
&& \hspace{0.9cm} \times  \left\{ \frac{ 32 \pi v^2 M^{}_0 \Delta M   }{ (c^{\prime 2}+2|e^\prime|^2) [256 \pi^2 v^4 \Delta M^2 + M^2_0 (d^{\prime 2}+2|f^\prime |^2)^2]} + (c^\prime \leftrightarrow d^\prime, e^\prime \leftrightarrow f^\prime) \right\}  \;,
\label{4.2.9}
\end{eqnarray}
where the second term in the brace differs from the first one by the interchanges $c^\prime \leftrightarrow d^\prime$ and $e^\prime \leftrightarrow f^\prime$, and $\varepsilon^{}_{\tau} \simeq - (1+ 2\Delta^{}_\tau) \varepsilon^{}_{\mu}$. Furthermore, one has
\begin{eqnarray}
\widetilde m^{}_\tau \simeq (1+ 2\Delta^{}_\tau) \widetilde m^{}_\mu =  \frac{1}{M^{}_0} (1+ 2\Delta^{}_\tau)  (|e^\prime|^2 +|f^\prime |^2 ) \;.
\label{4.2.10}
\end{eqnarray}
Consequently, the final baryon asymmetry is given by
\begin{eqnarray}
Y^{}_{\rm B} = c r \varepsilon^{}_{\mu} \left\{ \kappa ( \widetilde m^{}_\mu ) - (1+ 2\Delta^{}_\tau) \kappa \left[ (1+ 2\Delta^{}_\tau) \widetilde m^{}_\mu  \right]  \right\}  \;.
\label{4.2.11}
\end{eqnarray}
We see that unless $\kappa ( \widetilde m^{}_\alpha)$ is strictly inversely proportional to $\widetilde m^{}_\alpha$, which is not the case as can be seen from Eq.~(\ref{3.2.7}), $Y^{}_{\rm B}$ will become non-vanishing.

{\bf NO case:}
In the NO case with $\sigma =\pi/2$ and $P^{}_N = {\rm diag}({\rm i}, {\rm i})$, where one has $\cos z = \cos \theta$ and $\sin z = \sin \theta$ for the Casas-Ibarra parametrization, the combinations of $c^\prime$, $d^\prime$, $e^\prime$ and $f^\prime$ in Eq.~(\ref{4.2.9}) appear as
\begin{eqnarray}
&& c^\prime d^\prime + 2{\rm Re}(e^{\prime *} f^\prime) = \sqrt{M^{}_1 M^{}_2} (m^{}_2 -  m^{}_3) \cos  \theta \sin  \theta  \;, \hspace{1cm} {\rm Im}(e^{\prime *} f^\prime) = - \frac{1}{2} \sqrt{M^{}_1 M^{}_2} \sqrt{ m^{}_2 m^{}_3} c^{}_{12} c^{}_{13} \;, \nonumber \\
&& c^{\prime 2}+2|e^\prime|^2 = M^{}_1 (m^{}_2 \cos^2 \theta + m^{}_3 \sin^2 \theta) \;, \hspace{1cm} d^{\prime 2}+2|f^\prime |^2 = M^{}_2( m^{}_2 \sin^2 \theta + m^{}_3 \cos^2 \theta) \;.
\label{4.2.12}
\end{eqnarray}
On the other hand, $\widetilde m^{}_\mu$ turns out to be
\begin{eqnarray}
&& \widetilde m^{}_\mu \simeq  \frac{1}{2} (m^{}_2 c^2_{12} + m^{}_3) \simeq (0.027-0.028) \ {\rm eV}  \;.
\label{4.2.13}
\end{eqnarray}
In order to show the dependence of $Y^{}_{\rm B}$ on $\theta$ and $D$, in Fig.~2(a) we have plotted the contour lines of $Y^{}_{\rm B}/Y^0_{\rm B}$ on the $\theta$-$D$ plane. Here we have only shown the results for $\Delta M >0$, considering that the results for $\Delta M <0$ can be inferred from the fact that $Y^{}_{\rm B}$ keeps invariant under the transformations $\Delta M \to - \Delta M$ and $\theta \to \pi - \theta$ (see Eq.~(\ref{4.2.9})).
These results show that $Y^{}_{\rm B}$ has no chance to reach $Y^0_{\rm B}$. It is found that the maximally allowed value $Y^{\rm max}_{\rm B}$ of $Y^{}_{\rm B}$ can only reach 34\% of $Y^0_{\rm B}$. And $Y^{\rm max}_{\rm B}$ is achieved at $(\theta, D) \simeq (0.75\pi, 10^{-14})$ and $ (0.25\pi,  - 10^{-14})$, which can be easily understood from Eq.~(\ref{4.2.9}).

Then, we consider the case that two right-handed neutrinos are exactly degenerate at the flavor-symmetry scale. In this case, Eq.~(\ref{4.2.4}) gives that $\theta^{}_{\rm R}$ is independent of $\Lambda^{}_{\rm FS}/M^{}_0$. On the other hand, $\theta^{}_{\rm R}$ should be vanishing for $\Lambda^{}_{\rm FS}=M^{}_0$ (i.e., in the absence of the renormalization group evolution effect). One is thus left with the only possibility of $c d + 2{\rm Re}(e^* f) =0$ (i.e., two columns of $M^{}_{\rm D}(\Lambda^{}_{\rm FS})$ being orthogonal to each other) \cite{joaquim}. Consequently, to have a viable leptogenesis, both the orthogonal relation between two columns of $M^{}_{\rm D}(\Lambda^{}_{\rm FS})$ and the $\mu$-$\tau$ reflection symmetry need to be broken. In this way the final baryon asymmetry is suppressed by $\Delta^2_\tau$ and certainly unable to reproduce the observed value.

{\bf IO case:}
In the IO case with $\sigma -\rho =0$ and $P^{}_N = {\rm diag}(1, 1)$, where one also has $\cos z = \cos \theta$ and $\sin z = \sin \theta$ for the Casas-Ibarra parametrization, $\varepsilon^{}_{\mu}$ and $\widetilde m^{}_\mu$ become
\begin{eqnarray}
&& \varepsilon^{}_{\mu} \simeq (m^{}_1 -  m^{}_2) s^{}_{13} \sin 2 \theta \frac{ 16 \pi v^2 M^2_0 \Delta M   }{256 \pi^2 v^4 \Delta M^2 + m^2_0 M^4_0 }  \;, \nonumber \\
&& \widetilde m^{}_\mu \simeq  \frac{1}{2} (m^{}_1 s^2_{12} + m^{}_2 c^2_{12}) \simeq  (0.025-0.026) \ {\rm eV}  \;.
\label{4.2.14}
\end{eqnarray}
Due to the heavy cancellation between $m^{}_1$ and $m^{}_2$, the final baryon asymmetry is highly suppressed as shown in Fig.~2(b).

In the case that two right-handed neutrinos are exactly degenerate at the flavor-symmetry scale, for the same reason as in the NO case, the final baryon asymmetry is suppressed by $\Delta^2_\tau$ and unable to reproduce the observed value.

%%%%%%%%%%%%%%%%%%%%%% Figure 4-1 %%%%%%%%%%%%%%%%%%%%%%
\begin{figure*}[t]
\centering
\includegraphics[width=6in]{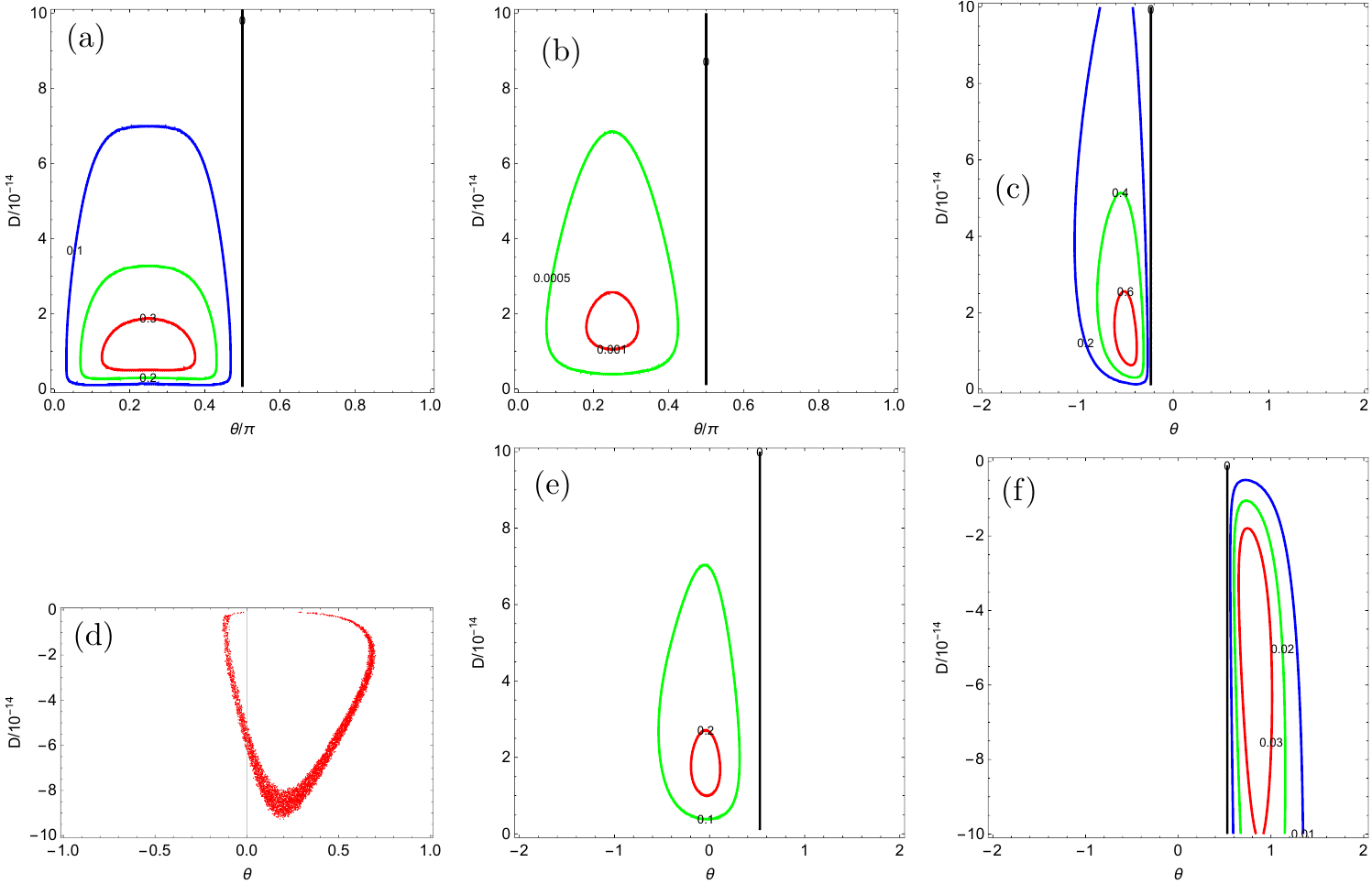}
\caption{ For the $\mu$-$\tau$ reflection symmetry scenario, the contour lines of $Y^{}_{\rm B}/Y^{0}_{\rm B}$ on the $\theta$-$D$ plane: the NO case with $P^{}_N = {\rm diag}({\rm i}, {\rm i})$ for $D>0$ (a); the IO case with $P^{}_N = {\rm diag}(1, 1)$ for $D>0$ (b); the NO case with $P^{}_N = {\rm diag}(1, {\rm i})$ for $D>0$ (c); the IO case with $P^{}_N = {\rm diag}(1, {\rm i})$ for $D>0$ (e) and $D<0$ (f). The values of $\theta$ and $D$ that allow $Y^{}_{\rm B}$ to reproduce $Y^{0}_{\rm B}$: the NO case with $P^{}_N = {\rm diag}(1, {\rm i})$ for $D<0$ (d). }
\label{Fig:4-1}
\end{figure*}
%%%%%%%%%%%%%%%%%%%%%%%%%%%%%%%%%%%%%%%%%%%%%%%%%%%

\subsubsection{$P^{}_N = {\rm diag}(1, {\rm i})$ or ${\rm diag}({\rm i}, 1)$}

In the case of $P^{}_N = {\rm diag}(1, {\rm i})$ or ${\rm diag}({\rm i}, 1)$, as a result of ${\rm Re}[(Y^\dagger_\nu Y^{}_\nu)^{}_{12}] = {\rm Re}\{ \pm {\rm i} [c d + 2{\rm Re}(e^* f)] \} =0$ (where the upper and lower signs of $\pm$ respectively correspond to $P^{}_N = {\rm diag}(1, {\rm i})$ and ${\rm diag}({\rm i}, 1)$, and similarly for $\mp$ in the following discussions) and $\Delta M^0_{21} \ll M^{}_0$, the right-handed neutrino mass matrix at the right-handed neutrino mass scale given by Eq.~(\ref{2.2.4}) approximates to
\begin{eqnarray}
M^{}_{\rm R} (M^{}_0) \simeq
\left( \begin{array}{cc}
M^{0}_1 - \displaystyle \frac{M^{}_0}{8\pi^2 v^2} ( c^2 + 2 |e|^2) \ln\left( \frac{\Lambda^{}_{\rm FS}}{M^{}_0} \right)  &  0  \cr
0 & M^{0}_2 - \displaystyle \frac{M^{}_0}{8\pi^2 v^2} ( d^2 + 2 |f|^2) \ln\left( \frac{\Lambda^{}_{\rm FS}}{M^{}_0} \right)  \cr
\end{array} \right) \;.
\label{4.2.15}
\end{eqnarray}
Consequently, the right-handed neutrino mass difference is given by
\begin{eqnarray}
\Delta M \simeq \Delta M^0_{21} + \frac{ M^{}_0} {8 \pi^2 v^2 } (c^2 + 2 |e|^2 - d^2 - 2 |f|^2 ) \ln\left( \displaystyle  \frac{\Lambda^{}_{\rm FS}}{M^{}_0} \right) \;.
\label{4.2.16}
\end{eqnarray} 
On the other hand, thanks to the renormalization group evolution effect of the Dirac neutrino mass matrix,
$\varepsilon^{}_{\alpha}$ become non-vanishing:
\begin{eqnarray}
&& \hspace{-1.cm} \varepsilon^{}_{e} \simeq -\Delta^{}_\tau cd  {\rm Im}(e^* f) \left\{ \frac{ 64 \pi v^2 M^{}_0 \Delta M   }{ (c^2+2|e|^2) [256 \pi^2 v^4 \Delta M^2 + M^2_0 (d^2+2|f|^2)^2]} + (c \leftrightarrow d, e \leftrightarrow f) \right\}  \;, \nonumber \\
&& \hspace{-1.cm} \varepsilon^{}_{\tau} \simeq \varepsilon^{}_{\mu} \simeq - \Delta^{}_\tau {\rm Re}(e^* f)  {\rm Im}(e^* f) \left\{ \frac{ 64 \pi v^2 M^{}_0 \Delta M   }{ (c^2+2|e|^2) [256 \pi^2 v^4 \Delta M^2 + M^2_0 (d^2+2|f|^2)^2]} + (c \leftrightarrow d, e \leftrightarrow f) \right\} \;.
\label{4.2.17}
\end{eqnarray}

{\bf NO case:}
In the NO case with $\sigma =0$ and $P^{}_N = {\rm diag}(1, {\rm i})$ (or ${\rm diag}({\rm i}, 1)$), where one has $\cos z = \cosh \theta$ and $\sin z = {\rm i} \sinh \theta $ (or $\cos z = {\rm i} \sinh \theta$ and $\sin z = \cosh \theta$) for the Casas-Ibarra parametrization, the combinations of $c$, $d$, $e$ and $f$ in Eqs.~(\ref{4.2.15}-\ref{4.2.17}) appear as
\begin{eqnarray}
&& c d \simeq \pm \sqrt{M^{}_1 M^{}_2} \sqrt{m^{}_2 m^{}_3} s^{}_{12} s^{}_{13} (\cosh^2 \theta + \sinh^2 \theta) + \sqrt{M^{}_1 M^{}_2} (m^{}_2 s^2_{12} + m^{}_3 s^2_{13}) \cosh \theta \sinh \theta  \;, \nonumber \\
&& {\rm Re}(e^* f) \simeq \frac{1}{2} \sqrt{M^{}_1 M^{}_2} \left[  m^{}_3  \cosh \theta \sinh \theta  \mp \sqrt{m^{}_2 m^{}_3} s^{}_{12} s^{}_{13} (\cosh^2 \theta + \sinh^2 \theta) \right] \;, \nonumber \\
&& c^2+2|e|^2 = M^{}_1( m^{}_2 |\cos z|^2 + m^{}_3 |\sin z|^2)  \;, \hspace{1cm}
d^2+2|f|^2 =  M^{}_2 (m^{}_2 |\sin z|^2 + m^{}_3 |\cos z|^2) \;,
\label{4.2.18}
\end{eqnarray}
while the expression of ${\rm Im}(e^* f)$ is same as that of ${\rm Im}(e^{\prime *} f^\prime)$ in Eq.~(\ref{4.2.12}).
And the washout mass parameters $\widetilde m^{}_\alpha$  are given by
\begin{eqnarray}
&& \widetilde m^{}_e \simeq (m^{}_2 s^2_{12} + m^{}_3 s^2_{13}) (\cosh^2 \theta + \sinh^2 \theta) \pm 4 \sqrt{m^{}_2 m^{}_3} s^{}_{12} s^{}_{13} \cosh \theta \sinh \theta   \;, \nonumber \\
&& \widetilde m^{}_\tau \simeq \widetilde m^{}_\mu \simeq \frac{1}{2} (m^{}_2 c^2_{12} + m^{}_3) (\cosh^2 \theta + \sinh^2 \theta)  \;.
\label{4.2.19}
\end{eqnarray}
Numerically, it is found that $Y^{}_{\rm B}$ has chance to reproduce $Y^0_{\rm B}$ for $D<0$. Figure~2(d) shows the values of $\theta$ and $D$ for realizing $Y^{}_{\rm B}= Y^0_{\rm B}$ within the $3\sigma$ level, while Fig.~2(c) shows the contour lines of $Y^{}_{\rm B}/Y^0_{\rm B}$ on the $\theta$-$D$ plane for $D>0$.
Here we have only shown the results for the case of $P^{}_N = {\rm diag}(1, {\rm i})$, while the results for the case of $P^{}_N ={\rm diag}({\rm i}, 1)$ can be inferred from the fact that $Y^{}_{\rm B}$ keeps invariant under the transformations $\Delta M \to - \Delta M$ and $\theta \to - \theta$ (see Eqs.~(\ref{4.2.17}-\ref{4.2.19})). From Fig.~2(c) and (d) we see that small values of $\theta$ (i.e., $|\theta| \lesssim 0.5$) are more favored by leptogenesis. This is because both the washout mass parameters $\widetilde m^{}_\alpha$ which suppress $\kappa(\widetilde m^{}_\alpha)$ and $c^2+2|e|^2$ and $d^2+2|f|^2$ which suppress $\varepsilon^{}_\alpha$ would increase rapidly for large values of $|\theta|$ (see Eqs.~(\ref{4.2.17}-\ref{4.2.19})).

In the case that two right-handed neutrinos are exactly degenerate at the flavor-symmetry scale, their mass difference induced by the renormalization group evolution effect is given by
\begin{eqnarray}
\Delta M \simeq  \pm \frac{1}{8\pi^2 v^2} (m^{}_2 - m^{}_3) M^2_0 \ln\left( \frac{\Lambda^{}_{\rm FS}}{M^{}_0} \right) \simeq \mp (0.14-0.16) \ {\rm eV} \;,
\label{4.2.192}
\end{eqnarray}
corresponding to $D \simeq \mp (1.4-1.6) \times 10^{-13}$.
For this value of $\Delta M$, the maximally allowed value $Y^{\rm max}_{\rm B}$ of $Y^{}_{\rm B}$ can only reach 66\% of $Y^0_{\rm B}$.

{\bf IO case:}
In the IO case with $\sigma -\rho = \pi/2$ and $P^{}_N = {\rm diag}(1, {\rm i})$ (or ${\rm diag}({\rm i}, 1)$), where one also has $\cos z = \cosh \theta$ and $\sin z = {\rm i} \sinh \theta $ (or $\cos z = {\rm i} \sinh \theta$ and $\sin z = \cosh \theta$) for the Casas-Ibarra parametrization, the combinations of $c$, $d$, $e$ and $f$ in Eq.~(\ref{4.2.17}) become
\begin{eqnarray}
&& c d \simeq \pm m^{}_0 \sqrt{M^{}_1 M^{}_2} c^{}_{12} s^{}_{12} (\cosh^2 \theta + \sinh^2 \theta) + m^{}_0 \sqrt{M^{}_1 M^{}_2} \cosh \theta \sinh \theta  \;, \nonumber \\
&& {\rm Re}(e^* f) \simeq \frac{1}{2}  m^{}_0  \sqrt{M^{}_1 M^{}_2} [\cosh \theta \sinh \theta \mp c^{}_{12}  s^{}_{12} (\cosh^2 \theta + \sinh^2 \theta) ] \;,  \nonumber \\
&&  {\rm Im}(e^* f) = - \frac{1}{2} m^{}_0 \sqrt{M^{}_1 M^{}_2} s^{}_{13} \;, \nonumber \\
&&  c^2+2|e|^2 \simeq  m^{}_0 M^{}_1 (\cosh^2 \theta + \sinh^2 \theta) \;, \hspace{1cm} d^2+2|f|^2 \simeq m^{}_0 M^{}_2 (\cosh^2 \theta + \sinh^2 \theta) \;.
\label{4.2.20}
\end{eqnarray}
And the washout mass parameters $\widetilde m^{}_\alpha$ turn out to be
\begin{eqnarray}
&& \widetilde m^{}_e \simeq m^{}_0 (\cosh^2 \theta + \sinh^2 \theta) \pm  4 m^{}_0 c^{}_{12} s^{}_{12} \cosh \theta \sinh \theta \;, \nonumber \\
&& \widetilde m^{}_\tau \simeq  \widetilde m^{}_\mu \simeq \frac{1}{2} m^{}_0 (\cosh^2 \theta + \sinh^2 \theta) \mp 2 m^{}_0 c^{}_{12} s^{}_{12} \cosh \theta \sinh \theta  \;.
\label{4.2.21}
\end{eqnarray}
Figure~2(e) and (f) (for $D>0$ and $D<0$, respectively) show the contour lines of $Y^{}_{\rm B}/Y^0_{\rm B}$ on the $\theta$-$D$ plane for the case of $P^{}_N = {\rm diag}(1, {\rm i})$. These results show that $Y^{}_{\rm B}$ has no chance to reach $Y^{0}_{\rm B}$, which can be understood from the great increase of $\widetilde m^{}_e$ compared to its counterpart in the NO case.

In the case that two right-handed neutrinos are exactly degenerate at the flavor-symmetry scale, their mass difference induced by the renormalization group evolution effect can be obtained from Eq.~(\ref{4.2.192}) by making the replacements $m^{}_2 \to m^{}_1$ and $m^{}_3 \to m^{}_2$. Owing to the heavy cancellation between $m^{}_1$ and $m^{}_2$, one arrives at $\Delta M \simeq \mp (2.3-2.8) \times 10^{-3}$ eV, corresponding to $D \simeq \mp (2.3-2.8) \times 10^{-15}$. For this value of $\Delta M$, the maximally allowed value $Y^{\rm max}_{\rm B}$ of $Y^{}_{\rm B}$ can only reach 8.2\% of $Y^0_{\rm B}$.

\section{High scale leptogenesis}

In this section, we consider the scenario that two right-handed neutrinos are not nearly degenerate any more. For this scenario, as mentioned in the introduction section, the requirement of successful leptogenesis places a lower bound $\sim 10^{9}$ GeV for the right-handed neutrino masses. Hence we just need to consider the unflavored and two-flavor regimes. On the other hand, the renormalization group evolution effect on the right-handed neutrino masses can be safely neglected: now in Eq.~(\ref{2.2.4}) the terms proportional to ${\rm ln}(\Lambda^{}_{\rm FS}/M^{}_0)$ are much smaller than $\Delta M^0_{21}$.

\subsection{Trimaximal mixing scenario}

For the Form Dominance realization of the trimaximal mixing, also due to the orthogonality relation between two columns of the Dirac neutrino mass matrix, the CP asymmetries for the decays of $N^{}_I$ are vanishing (see Eq.~(\ref{2.1.5})), prohibiting leptogenesis to proceed. As before, we study the possibility of the renormalization group evolution effect inducing a viable leptogenesis. For definiteness, in the following numerical calculations, we will take a benchmark value of $\Lambda^{}_{\rm FS}/M^{}_0=10$, corresponding to $\Delta^{}_\tau \simeq 2.36 \times 10^{-6}$.

In the unflavored regime, since the total CP asymmetry $\varepsilon^{}_I$ for the decays of $N^{}_I$ is proportional to ${\rm Im}\left[(M^\dagger_{\rm D} M^{}_{\rm D})^2_{IJ} \right]$ (see Eq.~(\ref{2.1.4})), only at the order of $\Delta^2_\tau$ can it become non-vanishing, rendering the final baryon asymmetry severely suppressed. For example, for an $M^{}_{\rm D}(\Lambda^{}_{\rm FS})$ of the form in Eq.~(\ref{3.1.5}) which gives a TM1 mixing in the IO case, the non-vanishing $\varepsilon^{}_I$ induced by the renormalization group evolution effect are given by
\begin{eqnarray}
\varepsilon^{}_{1} \simeq  \Delta^2_{\tau} \frac{m^{}_0 M^{}_2}{12 \pi v^2 I^{\prime 2}_0} \frac{{\rm Im}\left[(1+r)^2 e^{2{\rm i}\phi^{}_{21}}\right]}{3+2|r|^2} {\cal F} \left( \frac{M^2_2}{M^2_1} \right) \;, \nonumber \\
\varepsilon^{}_{2} \simeq  - \Delta^2_{\tau} \frac{m^{}_0 M^{}_1}{12 \pi v^2 I^{\prime 2}_0} \frac{{\rm Im}\left[(1+r)^2 e^{2{\rm i}\phi^{}_{21}}\right]}{3+2|r|^2} {\cal F} \left( \frac{M^2_1}{M^2_2} \right) \;.
\label{5.1.1}
\end{eqnarray}
Here $I^\prime_0$ is an analogue of $I^{}_0$ in Eq.~(\ref{2.2.3}) with the replacement $\ln (\Lambda^{}_{\rm FS}/M^{}_0) \to \ln (M^{}_0/\Lambda^{}_{\rm EW})$. For the typical value of $M^{}_0 = 10^{12}$ GeV, one has $I^\prime_0 \simeq 0.86$.
It arises because the leptogenesis scale $M^{}_0$ is far away from the electroweak scale $\Lambda^{}_{\rm EW}$ where the experimental data are utilized to infer the values of the model parameters.

{\bf TM1 mixing:} In the two-flavor regime, for an $M^{}_{\rm D}(\Lambda^{}_{\rm FS})$ of the form in Eq.~(\ref{3.1.5}), the non-vanishing $\varepsilon^{}_{I\alpha}$ induced by the renormalization group evolution effect appear as
\begin{eqnarray}
&& \varepsilon^{}_{1\tau} \simeq  \Delta^{}_{\tau} \frac{m^{}_0 M^{}_2}{24 \pi v^2 I^{\prime 2}_0} \frac{{\rm Im}\left[(1+r)^2 e^{2{\rm i}\phi^{}_{21}}\right]}{3+2|r|^2} {\cal F} \left( \frac{M^2_2}{M^2_1} \right) \;, \hspace{1cm} \varepsilon^{}_{1\gamma} \simeq - \varepsilon^{}_{1\tau} \;; \nonumber \\
&& \varepsilon^{}_{2\tau} \simeq  - \Delta^{}_{\tau} \frac{m^{}_0 M^{}_1}{24 \pi v^2 I^{\prime 2}_0} \frac{{\rm Im}\left[(1+r)^2 e^{2{\rm i}\phi^{}_{21}}\right]}{3+2|r|^2} {\cal F} \left( \frac{M^2_1}{M^2_2} \right) \;, \hspace{1cm} \varepsilon^{}_{2\gamma} \simeq - \varepsilon^{}_{2\tau} \;.
\label{5.1.2}
\end{eqnarray}
And the corresponding washout mass parameters are given by
\begin{eqnarray}
&& \widetilde m^{}_{1\tau} = \frac{1}{6 I^{\prime 2}_0} m^{}_1 \;, \hspace{1cm} \widetilde m^{}_{1\gamma} = \frac{5}{6 I^{\prime 2}_0 } m^{}_1 \;; \nonumber \\
&& \widetilde m^{}_{2\tau} = \frac{ |1+r|^2 }{(3+2 |r|^2) I^{\prime 2}_0 } m^{}_2 \;, \hspace{1cm} \widetilde m^{}_{2\gamma} = \frac{ 1+ |1-r|^2 }{(3+2 |r|^2) I^{\prime 2}_0 } m^{}_2 \;.
\label{5.1.3}
\end{eqnarray}
The final baryon asymmetry $Y^{}_{\rm B}$ can be calculated according to Eq.~(\ref{2.1.6}). We first consider the case of $M^{}_1 < M^{}_2$ where the contribution to leptogenesis mainly comes from $N^{}_1$. In this case, the resulting $Y^{}_{\rm B}/Y^0_{\rm B}$ is shown as a function of $\phi^{}_{21}$ in Fig~3(a). We see that the maximally allowed value of $Y^{}_{\rm B}/Y^0_{\rm B}$ can only reach $1.8 \times 10^{-5}$. In obtaining these results, we have taken the maximally allowed value of $M^{}_1$ (i.e., $10^{12}$ GeV) for the two-flavor regime to be viable as a benchmark value and allowed $M^{}_2$ to vary in the range $(2, 10)M^{}_1$ and the neutrino oscillation parameters in their $3\sigma$ ranges. Note that the results for smaller values of $M^{}_1$ can be obtained by scaling down the results in Fig~3(a) proportionally. This is because one has ${\cal F} \left( M^2_2/M^2_1 \right) \simeq -3 M^{}_1/(2 M^{}_2)$ for hierarchical $M^{}_{1}$ and $M^{}_2$, rendering $\varepsilon^{}_{1\tau}$ simply proportional to $M^{}_1$. For the case of $M^{}_1 < M^{}_2$ where the contribution to leptogenesis mainly comes from $N^{}_2$, similar results are obtained (see Fig.~3(b)). In obtaining these results, we have taken $M^{}_2 = 10^{12}$ GeV and allowed $M^{}_1$ to vary in the range $(2, 10)M^{}_2$ instead.

{\bf TM2 mixing:} For an $M^{}_{\rm D}(\Lambda^{}_{\rm FS})$ of the form in Eq.~(\ref{3.1.10}) which yields a TM2 mixing, in the NO case one has
\begin{eqnarray}
&& \varepsilon^{}_{1\tau} \simeq  \Delta^{}_{\tau} \frac{m^{}_3 M^{}_2}{24 \pi v^2 I^{\prime 2}_0 } \frac{{\rm Im}\left[(1+r)^2 e^{2{\rm i}\phi^{}_{21}}\right]}{1+3|r|^2} {\cal F} \left( \frac{M^2_2}{M^2_1} \right) \;, \hspace{1cm} \varepsilon^{}_{1\gamma} \simeq - \varepsilon^{}_{1\tau} \;; \nonumber \\
&& \varepsilon^{}_{2\tau} \simeq  - \Delta^{}_{\tau} \frac{m^{}_2 M^{}_1}{24 \pi v^2 I^{\prime 2}_0 } \frac{{\rm Im}\left[(1+r)^2 e^{2{\rm i}\phi^{}_{21}}\right]}{1+3|r|^2} {\cal F} \left( \frac{M^2_1}{M^2_2} \right) \;, \hspace{1cm} \varepsilon^{}_{2\gamma} \simeq - \varepsilon^{}_{2\tau} \;,
\label{5.1.4}
\end{eqnarray}
and
\begin{eqnarray}
&& \widetilde m^{}_{1\tau} = \frac{1}{3 I^{\prime 2}_0 } m^{}_2 \;, \hspace{1cm} \widetilde m^{}_{1\gamma} = \frac{2}{3 I^{\prime 2}_0 } m^{}_2 \;; \nonumber \\
&& \widetilde m^{}_{2\tau} = \frac{ |1+r|^2 }{2 (1+3 |r|^2) I^{\prime 2}_0 } m^{}_3 \;, \hspace{1cm} \widetilde m^{}_{2\gamma} = \frac{ 4|r|^2+ |1-r|^2 }{2(1+3 |r|^2) I^{\prime 2}_0 } m^{}_3 \;,
\label{5.1.5}
\end{eqnarray}
while the results in the IO case can be obtained by simply making the replacement $m^{}_3 \to m^{}_1$. The results in Fig.~3(c)--(f) show that the final baryon asymmetry is also smaller than the observed value by 5--6 orders of magnitude.

%%%%%%%%%%%%%%%%%%%%%% Figure 4-1 %%%%%%%%%%%%%%%%%%%%%%
\begin{figure*}[t]
\centering
\includegraphics[width=6in]{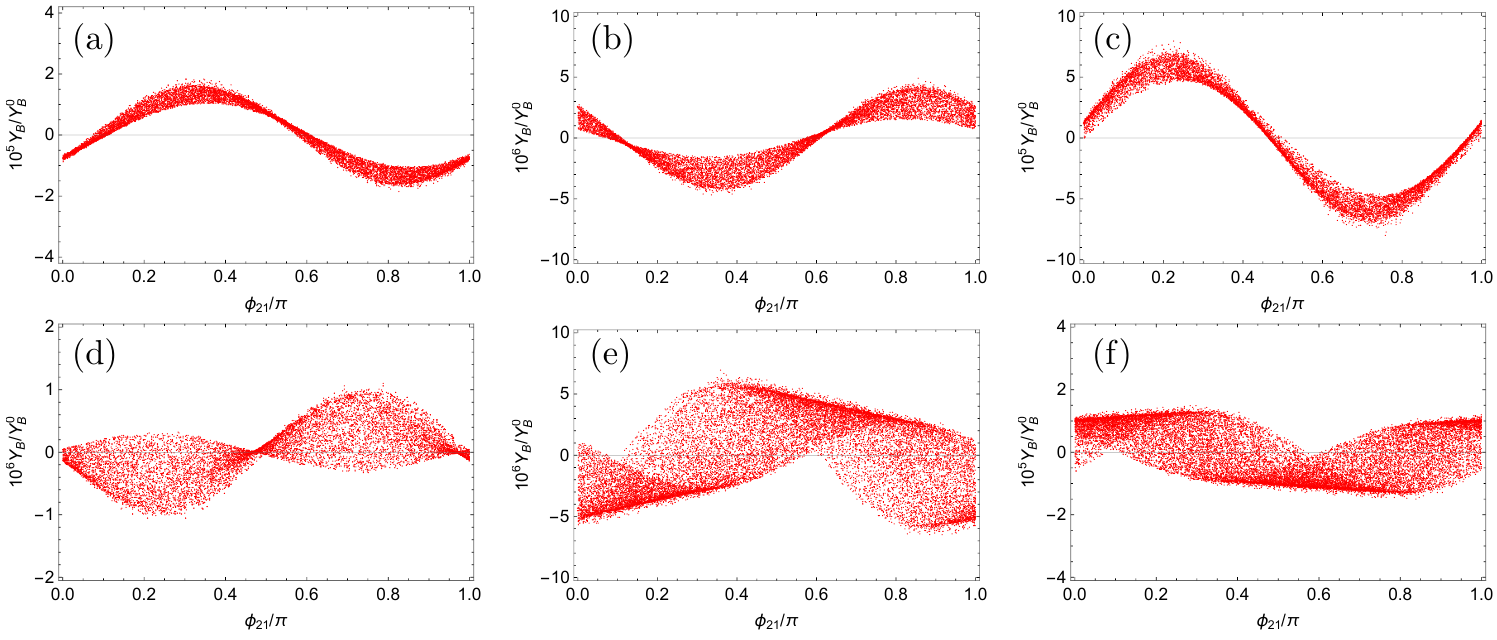}
\caption{ The rescaled $Y^{}_{\rm B}/Y^0_{\rm B}$ as functions of $\phi^{}_{21}$: the IO case with the TM1 mixing for $M^{}_1 < M^{}_2$ (a) and $M^{}_2 < M^{}_1$ (b); the NO case with the TM2 mixing for $M^{}_1 < M^{}_2$ (c) and $M^{}_2 < M^{}_1$ (d); the IO case with the TM2 mixing for $M^{}_1 < M^{}_2$ (e) and $M^{}_2 < M^{}_1$ (f). }
\label{Fig:5-1}
\end{figure*}
%%%%%%%%%%%%%%%%%%%%%%%%%%%%%%%%%%%%%%%%%%%%%%%%%%%

\subsection{Mu-tau reflection symmetry scenario}

For the mu-tau reflection symmetry scenario, Ref.~\cite{MN} shows that in the two-flavor regime leptogenesis has chance to be viable even when the symmetry keeps intact. In this regime, the renormalization group evolution effect only brings some tiny quantitative corrections but no qualitatively new features. Hence we just consider the unflavored regime where leptogenesis cannot proceed unless the mu-tau reflection symmetry is broken (by the renormalization group evolution effect here).

Let us first consider the case of $P^{}_N = {\rm diag}(1, 1)$ or ${\rm diag}({\rm i}, {\rm i})$.  In this case, the non-vanishing $\varepsilon^{}_{I}$ induced by the renormalization group evolution effect and the corresponding washout mass parameters are given by
\begin{eqnarray}
&& \varepsilon^{}_{1} \simeq - \Delta^{}_\tau \frac{cd + 2 {\rm Re}(e^* f)}{2\pi(c^2+2|e|^2) v^2} {\rm Im}(e^* f) {\cal F} \left( \frac{M^2_2}{M^2_1} \right) \;, \hspace{1cm} \widetilde m^{}_1 = \frac{1}{M^{}_1} ( c^2+2|e|^2 )  \;, \nonumber \\
&& \varepsilon^{}_{2} \simeq \Delta^{}_\tau \frac{cd + 2 {\rm Re}(e^* f)}{2\pi(d^2+2|f|^2) v^2} {\rm Im}(e^* f) {\cal F} \left( \frac{M^2_1}{M^2_2} \right) \;, \hspace{1cm}  \widetilde m^{}_2 = \frac{1}{M^{}_2} ( d^2+2|f|^2 ) \;.
\label{5.2.1}
\end{eqnarray}
In the NO case with $\sigma =\pi/2$ and $P^{}_N = {\rm diag}({\rm i}, {\rm i})$, where one has $\cos z = \cos \theta$ and $\sin z = \sin \theta$ for the Casas-Ibarra parametrization, the combinations of $c$, $d$, $e$ and $f$ in Eq.~(\ref{5.2.1}) can be obtained from their counterparts in Eq.~(\ref{4.2.12}) divided by $I^{\prime 2}_0$. In this case, the resulting $Y^{}_{\rm B}/Y^0_{\rm B}$ is shown as a function of $\theta$ in Fig.~4(a) and (b) (for $M^{}_1< M^{}_2$ and $M^{}_2 < M^{}_1$, respectively). We see that the maximally allowed value of $Y^{}_{\rm B}/Y^0_{\rm B}$ can only reach $4.0 \times 10^{-4}$. In obtaining these results, we have taken a benchmark value  $10^{13}$ GeV for the lighter right-handed neutrino mass. Note that one cannot enhance the final baryon asymmetry by increasing the right-handed neutrino masses at will, because the former would be exponentially suppressed by the $\Delta L=2$ lepton-number-violating processes mediated by the right-handed neutrinos if the latter were $\gtrsim 10^{14}$ GeV \cite{giudice}.
In the IO case with $\sigma - \rho =0$ and $P^{}_N = {\rm diag}(1, 1)$, where one also has $\cos z = \cos \theta$ and $\sin z = \sin \theta$ for the Casas-Ibarra parametrization, Eq.~(\ref{5.2.1}) turn out to be
\begin{eqnarray}
&& \varepsilon^{}_{1} \simeq  \Delta^{}_\tau \frac{M^{}_2 (m^{}_1 - m^{}_2) s^{}_{13} \sin 2 \theta}{8\pi v^2 I^{\prime 2}_0 } {\cal F} \left( \frac{M^2_2}{M^2_1} \right) \;, \hspace{1cm} \widetilde m^{}_1 \simeq \frac{m^{}_0}{I^{\prime 2}_0}  \;, \nonumber \\
&& \varepsilon^{}_{2} \simeq - \Delta^{}_\tau \frac{M^{}_1 (m^{}_1 - m^{}_2) s^{}_{13} \sin 2 \theta}{8\pi v^2 I^{\prime 2}_0} {\cal F} \left( \frac{M^2_1}{M^2_2} \right)  \hspace{1cm} \widetilde m^{}_2 \simeq \frac{m^{}_0}{I^{\prime 2}_0}  \;.
\label{5.2.2}
\end{eqnarray}
In this case, due to the heavy cancellation between $m^{}_1$ and $m^{}_2$, $\varepsilon^{}_{I}$ are much smaller than in the NO case.
The results of $Y^{}_{\rm B}$ for this case are shown in Fig.~4(c) and (d). We see that the final baryon asymmetry is smaller than the observed value by 7 orders of magnitude.

For the case of $P^{}_N = {\rm diag}(1, {\rm i})$ or ${\rm diag}({\rm i}, 1)$, the non-vanishing $\varepsilon^{}_{I}$ induced by the renormalization group evolution effect just differ by a sign with their counterparts in Eq.~(\ref{5.2.1}), while $\tilde m^{}_I$ are same as in Eq.~(\ref{5.2.1}).
In the NO case with $\sigma =0$ and $P^{}_N = {\rm diag}(1, {\rm i})$ (or ${\rm diag}({\rm i}, 1)$), where one has $\cos z = \cosh \theta$ and $\sin z = {\rm i} \sinh \theta $ (or $\cos z = {\rm i} \sinh \theta$ and $\sin z = \cosh \theta$) for the Casas-Ibarra parametrization, the combinations of $c$, $d$, $e$ and $f$ in Eq.~(\ref{5.2.1}) can be obtained from their counterparts in Eq.~(\ref{4.2.18}) divided by $I^{\prime 2}_0$. The results of $Y^{}_{\rm B}$ for this case are shown in Fig.~4(e) and (f). We see that the final baryon asymmetry is smaller than the observed value by 4--5 orders of magnitude.
In the IO case with $\sigma -\rho = \pi/2$ and $P^{}_N = {\rm diag}(1, {\rm i})$ (or ${\rm diag}({\rm i}, 1)$), where one also has $\cos z = \cosh \theta$ and $\sin z = {\rm i} \sinh \theta $ (or $\cos z = {\rm i} \sinh \theta$ and $\sin z = \cosh \theta$) for the Casas-Ibarra parametrization, $\varepsilon^{}_{I}$ and $\widetilde m^{}_I$ turn out to be
\begin{eqnarray}
&& \varepsilon^{}_{1} \simeq - \Delta^{}_\tau \frac{ m^{}_0 M^{}_2 s^{}_{13} \cosh \theta \sinh \theta }{ 2 \pi (\cosh^2 \theta + \sinh^2 \theta) v^2 I^{\prime 2}_0}  {\cal F} \left( \frac{M^2_2}{M^2_1} \right) \;, \hspace{1cm} \widetilde m^{}_1 =  m^{}_0 (\cosh^2 \theta + \sinh^2 \theta) \;, \nonumber \\
&& \varepsilon^{}_{2} \simeq  \Delta^{}_\tau \frac{ m^{}_0 M^{}_1 s^{}_{13} \cosh \theta \sinh \theta }{ 2 \pi (\cosh^2 \theta + \sinh^2 \theta) v^2 I^{\prime 2}_0}  \;, \hspace{1cm}  \widetilde m^{}_2 = m^{}_0 (\cosh^2 \theta + \sinh^2 \theta) \;.
\label{5.2.3}
\end{eqnarray}
The results of $Y^{}_{\rm B}$ for this case are shown in Fig.~4(g) and (h). We see that the final baryon asymmetry is smaller than the observed value by 5 orders of magnitude.

%%%%%%%%%%%%%%%%%%%%%% Figure 4-1 %%%%%%%%%%%%%%%%%%%%%%
\begin{figure*}[t]
\centering
\includegraphics[width=6in]{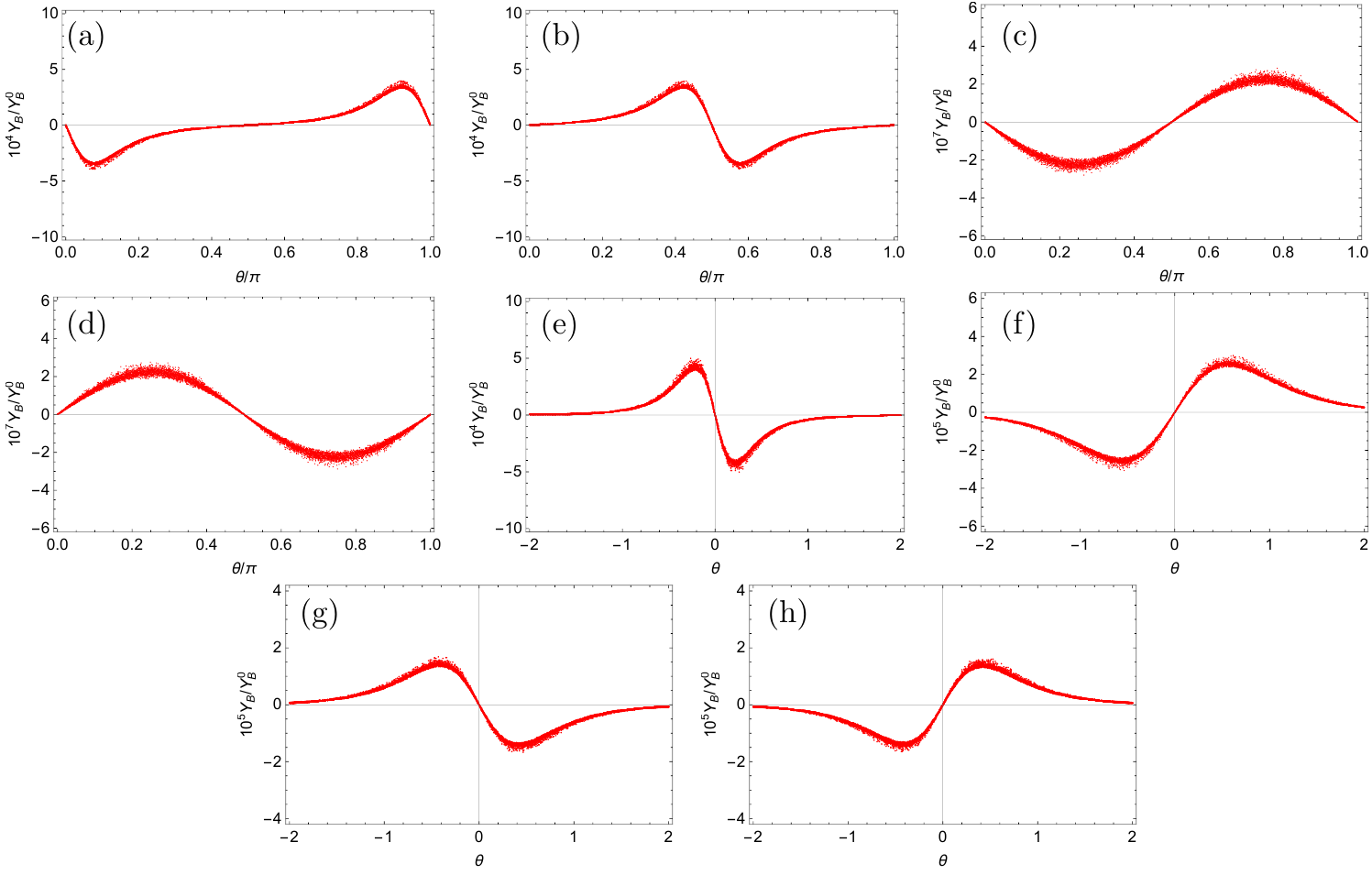}
\caption{ For the $\mu$-$\tau$ reflection symmetry scenario, the rescaled $Y^{}_{\rm B}/Y^0_{\rm B}$ as functions of $\theta$: the NO case with $P^{}_N = {\rm diag}({\rm i}, {\rm i})$ for $M^{}_1 < M^{}_2$ (a) and $M^{}_2 < M^{}_1$ (b); the IO case with $P^{}_N = {\rm diag}(1, 1)$ for $M^{}_1 < M^{}_2$ (c) and $M^{}_2 < M^{}_1$ (d); the NO case with $P^{}_N = {\rm diag}(1, {\rm i})$ for $M^{}_1 < M^{}_2$ (e) and $M^{}_2 < M^{}_1$ (f); the IO case with $P^{}_N = {\rm diag}(1, {\rm i})$ for $M^{}_1 < M^{}_2$ (g) and $M^{}_2 < M^{}_1$ (h). }
\label{Fig:5-2}
\end{figure*}
%%%%%%%%%%%%%%%%%%%%%%%%%%%%%%%%%%%%%%%%%%%%%%%%%%%

\section{Summary}

As we know, the most popular and natural way of generating the tiny but non-zero neutrino masses is the type-I seesaw mechanism, which also provides an appealing explanation for the baryon asymmetry of the Universe via the leptogenesis mechanism. But the conventional type-I seesaw model with three super heavy right-handed neutrino fields have the shortcomings that its parameters are many more than the low-energy neutrino parameters so that it lacks predictive power, and that the conventional seesaw scale is too high to be accessed by current experiments.

The experimental results that the neutrino mixing angles are close to some special values and $\delta$ is likely to be around $-\pi/2$ suggest that there may be some underlying flavor symmetry in the lepton sector. Along this direction, the trimaximal mixing and $\mu$-$\tau$ reflection symmetry are two typical examples. In this paper, we consider their imbedding in the minimal seesaw model (which contains much fewer parameters than the general seesaw model) with two TeV-scale right-handed neutrinos (for realizing a low-scale seesaw)  of nearly degenerate masses (for realizing a resonant leptogenesis).

Following the idea of Form Dominance (i.e., a specific right-handed neutrino is associated with a specific light neutrino mass eigenstate), the TM1 (TM2) mixing can be naturally achieved by having two columns of $M^{}_{\rm D}$ be respectively proportional and orthogonal to the first (second) column of the TBM mixing matrix (see Eqs.~(\ref{3.1.5}, \ref{3.1.10})).
However, due to the orthogonality relation between two columns of $M^{}_{\rm D}$ which leads the CP asymmetries $\varepsilon^{}_{\alpha}$ to be vanishing, leptogenesis cannot work.
Considering that the flavor symmetries which shape the special forms of $M^{}_{\rm D}$ are usually placed at a very high energy scale $\Lambda^{}_{\rm FS}$, one should take account of the renormalization group evolution effect when dealing with leptogenesis which takes place around the right-handed neutrino mass scale. Because of the difference among the charged-lepton Yukawa couplings, three rows of $M^{}_{\rm D}$ will receive different corrections (see Eq.~(\ref{2.2.2})). In this way, the orthogonality relation between two columns of $M^{}_{\rm D}$ will be broken, thus allowing leptogenesis to proceed. A detailed analysis shows that in the NO case with the TM2 mixing, the final baryon asymmetry $Y^{}_{\rm B}$ thus generated can successfully reproduce the observed value $Y^{0}_{\rm B}$. In comparison, in the IO case with either the TM1 or the TM2 mixing, the maximally allowed value of $Y^{}_{\rm B}$ can only reach 58\% or 54\% of $Y^{0}_{\rm B}$ (with $\Lambda^{}_{\rm FS} = 10^{10}$ GeV as a typical input). But note that for this case a viable leptogenesis can be easily achieved in the MSSM framework where the renormalization group evolution effect can be greatly enhanced by large $\tan \beta$ values.

In the presence of the $\mu$-$\tau$ reflection symmetry, the Dirac neutrino matrix will be restricted to a form as $M^{}_{\rm D} = M^{(0)}_{\rm D} P^{}_N$ where $M^{(0)}_{\rm D}$ is of the form in Eq.~(\ref{4.1.1}) and $P^{}_N = {\rm diag}(\eta^{}_1, \eta^{}_2)$ (for $\eta^{}_I = 1$ or i). For the case of $P^{}_N ={\rm diag} (1, 1)$ or ${\rm diag}({\rm i}, {\rm i})$, one has $\varepsilon^{}_{e} =0$ and $\varepsilon^{}_{\tau} = - \varepsilon^{}_{\mu}$ which in combination with $\widetilde m^{}_\mu = \widetilde m^{}_\tau $ lead $Y^{}_{\rm B}$ to be vanishing. For the case of $P^{}_N = {\rm diag} (1, {\rm i})$ or ${\rm diag} ({\rm i}, 1)$, one simply has $\varepsilon^{}_{\alpha} =0$, so leptogenesis cannot work either. When the renormalization group evolution effect is included, which will break the symmetry between the $\mu$ and $\tau$ flavors, leptogenesis also becomes possible. A detailed analysis shows that only in the NO case with $P^{}_N = {\rm diag}(1, {\rm i})$ (or ${\rm diag}({\rm i}, 1)$) can a viable leptogenesis be achieved. But in the other cases $Y^{0}_{\rm B}$ cannot be reproduced unless the MSSM is realistic.

For these scenarios, we have also considered the case that two right-handed neutrinos are exactly degenerate at the flavor-symmetry scale so that their mass difference is completely due to the renormalization group evolution effect. Unfortunately, in this case the final baryon asymmetry is unable to reproduce the observed value.

For completeness, we have also extended our analysis to the scenario that two right-handed neutrinos are not nearly degenerate any more. It turns out that the final baryon asymmetry is smaller than the observed value by several orders of magnitude.

\vspace{1cm}

\underline{Acknowledgments} \hspace{0.2cm}

We are deeply indebted to Professor Zhi-zhong Xing for bringing our attention to this interesting topic and Di Zhang and Shun Zhou for helpful discussions.
This work is supported in part by the National Natural Science Foundation of China under grant Nos.~11605081 and 12047570, and the Natural Science Foundation of the Liaoning Scientific Committee under grant NO.~2019-ZD-0473.

\end{document}